\documentclass[
  twocolumn,
]{aastex701}

\usepackage{
graphicx,
bm,
amsmath, 
amssymb,
mathtools,
tikz,
mathdots,
minted,
tabularx,
}
\usepackage[dvipsnames]{xcolor}
\usetikzlibrary{calc}

\newcommand{\entity}{\texttt{Entity}}

\newcommand{\iPhalf}{i+1/2}
\newcommand{\jPhalf}{j+1/2}

\newcommand{\iMhalf}{i-1/2}

\newcommand{\AtN}[2]{{}^{(\mathtt{#2})}{#1}}

\newcommand{\AtIJ}[3]{{}{#1_{(\mathtt{#2,#3})}}}
\newcommand{\CovAtIJ}[4]{{}{#1_{#2(\mathtt{#3,#4})}}}

\newcommand{\AtNIJ}[4]{{}^{(\mathtt{#2})}{#1_{(\mathtt{#3,#4})}}}
\newcommand{\CovAtNIJ}[5]{{}^{(\mathtt{#3})}{#1_{#2(\mathtt{#4,#5})}}}

\graphicspath{{./}{figures/}}

\begin{document}

\title{\entity~-- Hardware-agnostic Particle-in-Cell Code for Plasma Astrophysics.\\
  II: General Relativistic Module}

\correspondingauthor{Alisa Galishnikova}

\author[orcid=0000-0002-6300-3191]{Alisa Galishnikova} 
\affiliation{Center for Computational Astrophysics, Flatiron Institute, 162 Fifth Ave., New York, NY 10010, USA}
\email[show]{astro.alisag@gmail.com}

\author[orcid=0000-0001-8939-6862]{Hayk Hakobyan} 
\affiliation{Center for Computational Astrophysics, Flatiron Institute, 162 Fifth Ave., New York, NY 10010, USA}
\affiliation{Physics Department \& Columbia Astrophysics Laboratory, Columbia University, New York, NY 10027, USA}
\affiliation{Computational Sciences Department, Princeton Plasma Physics Laboratory (PPPL), Princeton, NJ 08540, USA}
\email{...}

\author[orcid=0000-0001-7801-0362]{Alexander Philippov}
\affiliation{Department of Physics, University of Maryland, College Park, MD 20742, USA}
\affiliation{Institute for Research in Electronics and Applied Physics, University of Maryland, College Park, MD 20742, USA}
\email{...}

\author[orcid=0000-0002-3241-9928]{Benjamin Crinquand}
\affiliation{IRAP, Université de Toulouse, CNRS, CNES, UPS, 14 avenue Edouard Belin, 31400 Toulouse, France}
\email{...}




\begin{abstract}
  Black hole environments often host plasmas that are fully collisionless or contain intrinsically collisionless regions, including relativistic jets and coronae where particle energization is ubiquitous. Capturing the physics of these systems requires numerical methods capable of modeling relativistic, magnetized, collisionless plasmas in strong gravitational fields. In this work, we introduce the general-relativistic module for the Entity -- the first open-source, coordinate-agnostic performance-portable particle-in-cell code. The code enables fast axisymmetric simulations of collisionless plasmas around black holes on any modern high-performance computing architecture (both GPUs and CPUs).
\end{abstract}


\section{Introduction}
\label{sec:intro}
Black holes (BHs) are fundamental astrophysical objects whose influence extends across a vast range of scales -- from stellar environments to the intergalactic medium -- where they play a central role in regulating the formation and evolution of cosmic structures. Because BHs cannot be observed directly, our understanding of them relies on interpreting their influence on the surrounding plasma, which typically consists of electrons, protons, heavier nuclei, and possibly positrons.

Significant progress in modeling such systems has been made through magnetohydrodynamic (MHD) simulations, which assume that the plasma is in local thermodynamical equilibrium. General relativistic MHD (GRMHD) further incorporates the effects of strong gravity, essential for accurately describing physics in curved spacetime \citep{EHTcode}. This framework has already yielded numerous insights into the global behavior of plasmas around BHs.

However, many phenomena associated with BHs remain poorly understood due to the numerical challenges involved in modeling the relevant plasma regimes. A prime example is the class of low-luminosity galactic nuclei, including M87* and Sgr A*, which have been imaged in unprecedented detail by the Event Horizon Telescope. These systems exhibit hot accretion flows \citep{YuanNarayan2014ARA&A..52..529Y}, where the plasma is essentially collisionless: the mean free path for Coulomb collisions exceeds the BH event horizon radius by about six orders of magnitude. Under such conditions, the plasma cannot easily thermalize, rendering the assumptions underlying MHD formally invalid. Nevertheless, MHD remains a valuable tool for capturing the large-scale dynamics governed by global conservation laws. Furthermore, additional justification for the fluid approximation can arise from effective ``fluidization'' -- that is, effective collisionality induced by small-scale plasma instabilities.

In addition to their quiescent emission, many such systems display episodic flaring activity characterized by non-thermal spectra. For instance, Sgr A* produces bright, multi-wavelength flares in the near-infrared, mid-infrared, and X-ray bands \citep{Dodds-Eden2009ApJ...698..676D, Yusef-Zadeh2009ApJ, GRAVITY2018, Fellenberg2025ApJ...979L..20V}. Understanding these flares requires a detailed investigation of the dynamics during such episodes, the mechanisms responsible for the generation of non-thermal particles, and their resulting radiation.

Other regions of interest that are fundamentally collisionless exist in both stellar-mass and supermassive BH systems. A notable example is relativistic jets, which, especially in the supermassive BH case, affect environments ranging from the immediate vicinity of the BH to the intracluster medium. Jet formation is widely attributed to the Blandford–Znajek (BZ) mechanism \citep{BlandfordZnajek1977MNRAS.179..433B}, but interpreting their morphology and emission requires modeling pair loading, particle acceleration, and radiation mechanisms -- processes beyond the scope of standard GRMHD.

Another example is the X-ray-emitting corona of efficiently accreting plasma flows, believed to be composed of a hot, magnetized, collisionless plasma located in the inner vicinity of the BH \citep{Galeev1979ApJ...229..318G, Beloborodov2017ApJ...850..141B, Bambic2024MNRAS.527.2895B}. The corona likely dissipates magnetic energy through combination of magnetic reconnection and turbulence.

These considerations motivate the need for a framework that treats the plasma as collisionless while still incorporating the effects of strong gravity. Such an approach is realized through the general-relativistic particle-in-cell (GRPIC) method, first developed by \citet{Parfrey2019PhRvL}. More recently, several new GRPIC codes have been introduced \citep{APERTURE, FPIC2025ApJ...992L...8M}. In this work, we present the first architecture-agnostic GRPIC code, which is publicly available as part of the \entity\ framework.\footnote{GitHub link: \href{https://github.com/entity-toolkit/entity}{github.com/entity-toolkit/entity}.}

The paper is structured as follows. In Section~\ref{sec:theory}, we introduce the general equations in the 3+1 formalism and define our notation. Section \ref{sec:methodology} describes the numerical algorithm implemented within the \entity\ framework. Finally, in Section~\ref{sec:tests}, we present a suite of test cases that validate our GRPIC implementation.

\section{Equations in curvilinear space-time}\label{sec:theory}
Throughout this paper, we use the following conventions: $c$ denotes the speed of light and is taken to be $1$ throughout this paper, $G$ the gravitational constant, and $M$ the mass of the central BH. We additionally define the gravitational radius as $r_g=GM$.

The evolution of electromagnetic fields and charged particles in curved spacetime is naturally described using the $3+1$ formalism. This approach foliates a Lorentzian manifold into a family of spatial hypersurfaces indexed by a global time coordinate. Extensive treatments of the formalism can be found in \citet{Komissarov_2004, ThorneMacDonald1982, DodinFisch2010PhPl, Gourgoulhon2012LNP}.

\subsection{Spacetime metric}
We adopt the $3+1$ decomposition of spacetime, where the manifold is endowed with a symmetric, non-degenerate metric tensor $g_{\mu \nu}$ of signature $(-+++)$, and restrict our analysis to stationary spacetimes. Throughout this work, Greek indices ($\mu, \nu, \dots$) run from 0 to 3, while Latin indices ($i, j, \dots$) range from 1 to 3 and denote spatial components.

A key component of the $3+1$ formalism is the introduction of a timelike unit normal vector to the spatial hypersurfaces, which we define as
\begin{equation}
  n_\mu = (-\alpha, 0, 0, 0),
\end{equation}
where $\alpha$ is the \emph{lapse function}, encoding the proper time separation between hypersurfaces. This vector represents the covariant four-velocity of a local fiducial observer (FIDO). The \emph{spatial metric} on each hypersurface -- also referred to as the projection tensor -- is given by
\begin{equation}
  h_{\mu \nu} = g_{\mu \nu} + n_\mu n_\nu,
\end{equation}
and has a Euclidean signature $(+++)$. It defines distances and inner products on the three-dimensional spatial slices.

The contravariant form of the FIDO four-velocity is
\begin{equation}
  n^\mu = \alpha^{-1}\begin{pmatrix}1\\\beta^i\end{pmatrix},
\end{equation}
where $\beta^i$ is the \emph{shift vector}, which describes the relative motion of the spatial coordinates with respect to the FIDO frame. The shift vector captures the non-orthogonality between the time coordinate and the spatial hypersurfaces. A useful identity relates the determinants of the full spacetime and spatial metrics:
\begin{equation}
  g \equiv \det(g_{\mu \nu}) = -\alpha^2 h, ~\text{where}~ h \equiv \det(h_{ij}).
\end{equation}

\noindent With these definitions, the spacetime interval in $3+1$ formalism reads
\begin{equation}
  ds^2 = (\beta^2 - \alpha^2)\, dt^2 + 2\beta_i\, dx^i dt + h_{ij}\, dx^i dx^j,
\end{equation}

\noindent where $\beta^2 = \beta_i \beta^i$, $t$ is the coordinate time, and $x^i$ denote the spatial coordinates.

Up to this point, all expressions assumed a general foliation of a Lorentzian manifold. In the case of a rotating Kerr BH, various foliations (and thus coordinate systems) are possible; in this work, we adopt \emph{Kerr--Schild coordinates} to describe rotating Kerr black holes. This coordinate system is widely used in numerical simulations of accretion flows due to its regularity at the event horizon (while other coordinates, such as the Boyer--Lindquist system, introduce coordinate singularities near the horizon). Details of the Kerr--Schild metric, including its components, derivatives, and coordinate transformations, are provided in Appendix~\ref{ap:metric}.

\subsection{Electromagnetic fields \& Maxwell's equations}

The covariant Maxwell tensor is defined as
\begin{equation}
  F_{\mu\nu} \equiv g_{\mu\alpha} g_{\nu\beta} F^{\alpha\beta} = \partial_\mu A_\nu - \partial_\nu A_\mu,
\end{equation}
where $A_\mu$ is the electromagnetic four-potential. Where for visual clarity we employ $\partial_\mu\equiv\partial/\partial x^\mu$. The dual Maxwell tensor, often referred to as the Faraday tensor, is given by
\begin{equation}
  {}^*F^{\alpha\beta} = \frac{1}{2} e^{\alpha \beta \mu \nu} F_{\mu \nu},
\end{equation}
where $e^{\alpha \beta \mu \nu} \equiv \varepsilon^{\alpha \beta \mu \nu} / \sqrt{-g}$, and $\varepsilon^{\alpha \beta \mu \nu}$ is the totally antisymmetric Levi-Civita symbol in four dimensions. In three spatial dimensions, we also define $e^{ijk} \equiv \varepsilon^{ijk}/\sqrt{h}$, where $\varepsilon^{ijk}$ is the 3D Levi-Civita symbol. For covariant components, the following identities hold:
\begin{equation*}
  e_{\alpha \beta \mu \nu} = \sqrt{-g} \, \varepsilon_{\alpha \beta \mu \nu}, \quad e_{ijk} = \sqrt{h} \, \varepsilon_{ijk}.
\end{equation*}

Maxwell's equations in covariant form are:
\begin{equation}
  \begin{aligned}
    \nabla_\beta {}^*F^{\alpha\beta} & = 0,              \\
    \nabla_\beta F^{\alpha\beta}     & = 4 \pi I^\alpha,
  \end{aligned}
\end{equation}
where $I^\alpha = (\rho, \bm{J})^T$ is the four-current density, containing the charge density $\rho$ and the spatial current density $\bm{J}$. To obtain Maxwell's equations in a more familiar form, we define the spatial components of the electromagnetic field as:
\begin{equation}
  \begin{aligned}
    B^i & \equiv \alpha \, {}^*F^{i0},                                                   \\
    E^i & = h^{ij} E_j,~\text{where}~ E_i \equiv \frac{\alpha}{2} e_{ijk} \, {}^*F^{jk},
  \end{aligned}
\end{equation}
which allows us to recover the familiar special-relativistic forms of the Faraday's law and the absence of magnetic monopoles from $\nabla_\beta {}^*F^{\alpha\beta} = 0$:
\begin{align}
  \nabla \cdot \bm{B}                   & = 0,                                             \\
  \frac{\partial \bm{B}}{\, \partial t} & = -\nabla \times \bm{E},\label{eq:maxwell-set-1}
\end{align}
where bold symbols denote contravariant 3-vectors, e.g., $\bm{B} \equiv B^i$. From these, we also find:
\begin{equation}
  \begin{aligned}
    E_i & \equiv F_{i0},                 \\
    B^i & = \frac{1}{2} e^{ijk} F_{jk} =
    \frac{\varepsilon^{ijk}}{\sqrt{h}} \frac{\partial A_k}{\partial x^j}.
  \end{aligned}\label{eq:EB-set}
\end{equation}

\noindent To express the second set of Maxwell's equations in the $3+1$ form, we additionally define:
\begin{equation}
  \begin{aligned}
    D^i & \equiv \alpha F^{i0},                                                    \\
    H^i & = h^{ij} H_j, ~\text{where}~ H_i \equiv \frac{\alpha}{2} e_{ijk} F^{jk}.
  \end{aligned}
\end{equation}

\noindent These definitions yield Gauss' and Ampr\`e's laws:
\begin{align}
  \nabla \cdot \bm{D}                & = 4\pi \rho,                                                   \\
  \frac{\partial \bm{D}}{\partial t} & = \nabla \times \bm{H} - 4\pi \bm{J}, \label{eq:maxwell-set-2}
\end{align}
from which charge conservation naturally follows.

By construction, the vectors $B^\mu$ and $D^\mu$ are spatial; they lie within each spatial hypersurface and are orthogonal to the normal vector $n_\mu$, satisfying:
\begin{equation}
  \begin{aligned}
    B^\mu & = -{}^*F^{\mu \nu} n_\nu, \\
    D^\mu & = F^{\mu \nu} n_\nu.
  \end{aligned}
\end{equation}

\noindent Thus, $\bm{B}$ and $\bm{D}$ are treated as the primary fields, while $\bm{E}$ and $\bm{H}$ are considered auxiliary fields. The relationships between them is given by the following equations:
\begin{equation}
  \begin{aligned}
    \bm{E} & = \alpha \, \bm{D} + \boldsymbol{\beta} \times \bm{B}, \\
    \bm{H} & = \alpha \, \bm{B} - \boldsymbol{\beta} \times \bm{D},
  \end{aligned}
\end{equation}
where $\alpha$ is the lapse function, and $\boldsymbol{\beta}$ is the shift vector in the $3+1$ decomposition of spacetime. In what follows, we simplify all our expressions assuming \(\beta^i = (\beta^1, 0, 0)^T\), which is appropriate for the Kerr--Schild coordinate system. These relations, along with the spatial metric $h^{ij}$, encapsulate all gravitational effects in the 3+1 formulation of Maxwell's equations. In the special relativistic limit where $\alpha = 1$ and $\boldsymbol{\beta} = 0$, these reduce to the familiar flat spacetime electromagnetic fields:
\[
  \bm{E} = \bm{D}, \quad \bm{H} = \bm{B}.
\]

\subsection{Particles}
In Lagrangian formulation, the equation of motion for a particle of mass \( m \) and charge \( q \), moving in a curved spacetime under the influence of an external electromagnetic field, is given by:
\begin{equation}
  \frac{d^2 x^\mu}{d\tau^2} + \Gamma^\mu_{\lambda \sigma} \frac{d x^\lambda}{d\tau} \frac{d x^\sigma}{d\tau} = \frac{q}{m} g_{\nu \rho} F^{\mu \nu} \frac{d x^\rho}{d\tau},
\end{equation}
where $x^\mu$ is the coordinate of the particle, $\tau$ is the particle's proper time, and $\Gamma^\mu_{\lambda \sigma}$ are the Christoffel symbols \citep{Weinberg_1972}. The left-hand side describes the geodesic motion of a free particle in curved spacetime, accounting for gravitational effects via the Christoffel symbols. The right-hand side represents the Lorentz force acting on the particle, generalized to curved spacetime using the covariant Maxwell tensor $F^{\mu\nu}$.

Using 3+1 decomposition of the Maxwell tensor and Hamiltonian formulation, these equations can be written in the following form:
\begin{equation}
  \begin{aligned}
    \frac{d x^i}{d t} & =\frac{u^i}{u^0} = \frac{\alpha}{\gamma} h^{ij} u_j - \beta^i, \\
    \frac{d u_i}{d t} & =
    \underbrace{
      -\gamma \partial_i \alpha + u_j\partial_i \beta^j - \frac{\alpha}{2\gamma} u_j u_k\partial_i h^{jk}
    }_{\text{``curvature'' force}}                                                     \\
                      & + \underbrace{
      \frac{q}{m}\alpha\left(h_{ij}D^j + \frac{1}{\sqrt{h}\gamma}\varepsilon_{ijk}h^{jl} u_l B^k\right),
    }_{\text{Lorentz force ($m\ne 0$)}}
    \label{eq:equation-of-motion}
  \end{aligned}
\end{equation}
where $t \equiv x^0$ is the coordinate time, $(u_0,u_i)$ is the particle's covariant four-velocity\footnote{The contravariant components can be obtained as \(u^i = h^{ij} u_j - u_0 \beta^i\), and the time component satisfies \(\alpha u^0 = \gamma\).}, and the FIDO-measured dimensionless energy, $\gamma$, can be found as:
\begin{align}
  \gamma =
  \begin{cases}
    \sqrt{1 + h^{ij} u_i u_j}, & \text{for massive particles},  \\
    \sqrt{h^{ij} u_i u_j},     & \text{for massless particles}.
  \end{cases}
\end{align}

Additionally, in the Hamiltonian formulation, two useful conservation laws naturally arise. The conjugate four-momentum of a charged particle is defined as
\begin{equation}
  \pi_\mu = u_{\mu} + \frac{q}{m} A_\mu,
\end{equation}
where \(u_\mu\) is the four-velocity, \(A_\mu\) is the electromagnetic four-potential, and \(p_{\mu} = m u_{\mu}\) is the mechanical four-momentum.
In spacetimes that admit time-translation and azimuthal symmetries, the corresponding conserved quantities are the total energy \(\mathcal{E}\) and the angular momentum \(\mathcal{L}\), given by
\begin{align}
  \frac{\mathcal{E}}{m} & = -\pi_0 = -u_0 - q A_0, \\
  \frac{\mathcal{L}}{m} & = \pi_3 = u_3 + q A_3.
\end{align}

\section{Particle-in-cell Algorithm In Curvilinear Space}
\label{sec:methodology}

\entity~ solves the Vlasov--Maxwell system of equations in flexible coordinates, by integrating the time-dependent Maxwell's Eqs.~\eqref{eq:maxwell-set-1} and \eqref{eq:maxwell-set-2} sourced by the currents imposed by macroparticles, which obey the equations of motion given by \eqref{eq:equation-of-motion}. Here, we describe the numerical algorithms and their implementation specific to the general relativistic module of the code. The current version supports 2.5D axisymmetric simulations in spherical coordinates, with a strong emphasis on simulations of black hole magnetospheres and accretion. Full 3D simulations using the cubed-sphere grid will be presented in a future publication.

In what follows, we adopt code units, following the notation described in paper I. The physical coordinates are mapped onto code coordinates such that the grid is uniform in the code space. Consequently, the code coordinates span a range from zero to the total number of grid cells in each dimension, {$\mathtt{N_\theta}$ and $\mathtt{N_r}$ for the radial and azimuthal axis, respectively.} Any offset from the grid nodes is represented by the displacement from the corner of the cell, denoted as $dx$. The position of each particle is stored using two quantities: the index of the cell it occupies and the displacement within that cell (e.g., \texttt{i1} and \texttt{dx1}); see paper I for details of the coordinate implementation. The components of the spatial metric $h_{ij}$ are expressed in normalized code units (see Appendix~\ref{ap:metric}). Unlike the special-relativistic (SR) module, GR assumes that the magnetic and electric field components supplied by the problem generator are already given in code units (the spatial staggering of each component is still done by the code automatically). This convention is convenient, as field quantities are often initialized using combinations of metric components.

We employ a finite-difference time-domain (FDTD) leap-frog scheme, discretizing both the electric $D^i$ and the magnetic $B^i$ fields in both time and space. As in SR, electric field components and particle positions are defined at integer timesteps, $\AtN{t}{n}$, while magnetic fields and particle four-velocities are staggered by half a timestep back, $\AtN{t}{n-1/2}$. Typewriter font (e.g., $\mathtt{i}$, $\mathtt{n}$) denotes indices on the discretized numerical grid in both time and space. The timestep is fixed, $\Delta t\equiv \AtN{t}{n} - \AtN{t}{n-1}$, such that $ \AtN{t}{n}\equiv \mathtt{n}\Delta t$.

Implementation in GR largely follows the SR structure. However, because $\bm{D}$ and $\bm{B}$ fields are coupled via the auxiliary fields $\bm{E}$ and $\bm{H}$, the leapfrog algorithm requires modifications: copies of fields and electric current at two different times are stored (more details to follow in Section~\ref{sec:pic_loop}). Some field components must be interpolated to grid positions not defined on the Yee mesh (see Section~\ref{sec:spatial-disc}), and the electric current deposition requires storing both the initial and the final particle positions (Section~\ref{sec:currents}).

\subsection{Particles}
\label{sec:particle-push}

We solve the equations of motion for macroparticles given in \eqref{eq:equation-of-motion}. Particle coordinates and velocities are staggered in time using a second-order accurate leapfrog scheme. In this approach, the velocity is advanced first and then used to update the position. At timestep $\mathtt{n}$, the velocity is updated from $\AtN{u_i}{n-1/2}$ to $\AtN{u_i}{n+1/2}$, after which the position is advanced from $\AtN{x^i}{n}$ to $\AtN{x^i}{n+1}$:
\begin{align}
  \begin{aligned}
    \AtN{u_i}{n-1/2} & \xRightarrow[\Delta t]{} \AtN{u_i}{n+1/2}, \\
    \AtN{x^i}{n}     & \xRightarrow[\Delta t]{} \AtN{x^i}{n+1}.
  \end{aligned}
\end{align}

\noindent The velocity update is performed using symmetric second-order accurate Strang splitting \citep{Parfrey2019PhRvL}:
\begin{equation}
  f(t + \Delta t) = \mathcal{O}_{\Delta t} f(t) \approx \mathcal{O}_{\Delta t/2}^{EM} \mathcal{O}_{\Delta t}^{GR} \mathcal{O}_{\Delta t/2}^{EM} f(t),
\end{equation}
where $\mathcal{O}_{\Delta t}^{GR}$ and $\mathcal{O}_{\Delta t}^{EM}$ denote the pure geodesic and electromagnetic update operators, respectively. These correspond to the “curvature” and Lorentz force terms in \eqref{eq:equation-of-motion}.

The electromagnetic stage of the particle pusher implemented in the GRPIC module follows the standard Boris algorithm, as described in paper I, which is widely employed in flat space-time particle-in-cell codes for its numerical stability and excellent long-term behavior. The principal modification for the GR case is the inclusion of a tetrad transformation that maps the particle velocity and electromagnetic field components from the global coordinate basis to a locally flat orthonormal frame. This transformation is applied at the beginning of each velocity update half-step in $\mathcal{O}_{\Delta t/2}^{EM}$, enabling the use of the SR Boris procedure in the local frame. After the push, the updated velocities are transformed back to the global coordinate basis using the inverse tetrads.

For the geodesic update, we employ an iterative midpoint integrator to advance both the velocity and the position of the particle. In this scheme, the new velocity (or position) is first predicted using an initial guess and then iteratively corrected based on the updated local metric quantities. For the geodesic velocity update, let $\AtN{u_i}{n-1/2}$ denote the components of the covariant four-velocity of the particle before the update. The initial guess (zeroth iteration) for the advanced velocity is taken as $\tilde{u}^{(0)}_i = \AtN{u_i}{n-1/2}$, and the solution is iteratively refined according to:

\begin{align}
  \begin{aligned}
    \tilde{u}^{(m+1)}_i    & = \AtN{u_i}{n-1/2} + \Delta t f_u^{GR}\left(\bar{u}_i\right),          \\
    \text{where}~\bar{u}_i & \equiv \frac{1}{2} \left( \AtN{u_i}{n-1/2} + \tilde{u}^{(m)}_i\right),
  \end{aligned}
\end{align}
\noindent where the solution approaches the exact value, $\tilde{u}^{(m)}_i\to\AtN{u_i}{n+1/2}$, as the number of iterations, $m$, grows. In practice, convergence of the iterative scheme is typically achieved within ten iterations; the fixed number of iterations, $\mathtt{n_{iter}}$, is a user-configurable parameter. The explicit form of the geodesic velocity update during each iteration is given with the following relation

\begin{equation}
  \label{eq:discretized-geodesic-u}
  \begin{aligned}
     & \frac{\tilde{u}^{(m+1)}_i- \AtN{u_i}{n-1/2}}{\Delta t} =
    - \alpha \bar{u}^0 \partial_i \alpha
    + \bar{u}_1 \partial_i \beta^1
    \\
     & \quad - \frac{1}{2 \bar{u}^0} \Bigg(
    \sum_{j=1}^3
    \partial_i h_{jj}\bar{u}_j^2
    + 2 \partial_i h_{13} \bar{u}_1  \bar{u}_3
    \Bigg),
    ~\text{for}~i=\{1,2\},                                      \\
     & \text{and}~\tilde{u}^{(m+1)}_3 = \AtN{u_3}{n}.
  \end{aligned}
\end{equation}

\noindent Here, $\bar{u}^0 = \bar{\gamma}/\alpha$, and $\bar{\gamma}$ is the FIDO-measured energy associated with the four-velocity $\bar{u}_i$. The exact analytic expressions for the derivatives of the metric components are used in the integration and are provided in Appendix~\ref{ap:metric}. Since the GR-push is performed after the half of the electromagnetic push (which is described in detail below), for the massive charged particles, this initial velocity guess in the former corresponds to that obtained after the first electromagnetic half-push, $\tilde{u}_i^{(0)} = \AtN{u^{\ast}_i}{n-1/2}$, while for the chargeless particles the EM push is skipped altogether. The velocity update involves coordinate-dependent metric coefficients computed at the position of the particle. Because our algorithm implements a leap-frog scheme, we use the position from the previous timestep, $\AtN{x^i}{n}$, for the metric coefficients in \eqref{eq:discretized-geodesic-u}.

The newly obtained velocity may then be used to update the particle's position, which we also perform iteratively. Given the previous timestep position, we take the initial guess as $\tilde{x}^{i(0)} = \AtN{x^i}{n}$, and iterate according to:
\begin{equation}
  \begin{aligned}
    \tilde{x}^{i(m+1)}     & = \AtN{x^i}{n} + \Delta t f_x^{GR}\left( \bar{x}^i \right),  \\
    \text{where}~\bar{x}^i & = \frac{1}{2} \left( \AtN{x^i}{n} + \tilde{x}^{i(m)}\right).
  \end{aligned}
\end{equation}
\noindent The full expression according Eq.~\eqref{eq:equation-of-motion} takes the following explicit form:
\begin{equation}
  \label{eq:discretized-geodesic-x}
  \begin{aligned}
    \frac{\tilde{x}^{i(m+1)} - \AtN{x^i}{n}}{\Delta t} & =
    \frac{h^{ij} \ \AtN{u_j}{n+1/2}}{u^0} - \beta^i,~\text{for}~i=\{1,2\}.
  \end{aligned}
\end{equation}

\noindent In Eq.~\eqref{eq:discretized-geodesic-x}, all the metric coefficients, as well as the $\alpha$ and $\beta^i$ are computed at an intermediate position, $\bar{x}^i$.


Although our algorithm does not strictly require it, we also record the out-of-plane angular coordinate, $x^3\equiv \phi$, for each of particles. This quantity is mainly used to test the particle pusher, as described in Sec.~\ref{sec:tests}. After each push, the coordinate is updated based on the newly computed velocity according to
\begin{equation}
  \label{eq:discretized-geodesic-x-phi}
  \begin{aligned}
    \frac{\AtN{x^3}{n+1} - \AtN{x^3}{n}}{\Delta t} & =
    \frac{h^{3j} \ \AtN{u_j}{n+1/2}}{u^0},
  \end{aligned}
\end{equation}

Below we describe the full numerical scheme for a single particle update at a timestep $\mathtt{n}$.
\begin{enumerate}
  \item Save previous location:
  \item[] $\AtN{x^i_{\rm prev}}{n+1} = \AtN{x^i}{n}$.

  \item Interpolate field components to the particle's initial location:
  \item[]  $D_p^i$, $B_p^i$.

  \item Transform the interpolated fields as well as the particle's velocity to the tetrad basis:
  \item[] $D_p^i, B_p^i \xRightarrow[]{\AtN{x^i}{n}} D_p^{\hat i}, B_p^{\hat i}$,
  \item[] $\AtN{u_i}{n-1/2} \xRightarrow[]{\AtN{x^i}{n}} \AtN{u_{\hat i}}{n-1/2}$.

  \item Perform the first EM push (in tetrad basis) for \(\Delta t/2\):
  \item[] $\AtN{u_{\hat i}}{n-1/2} \xRightarrow[\Delta t/2]{D_p^{\hat i}, B_p^{\hat i}} \AtN{u^{\ast}_{\hat i}}{n-1/2} $.

  \item Transform the partially-updated velocity back to the covariant basis at $\AtN{x^i}{n}$:
  \item[] $\AtN{u^{\ast}_{\hat i}}{n-1/2} \xRightarrow[]{\AtN{x^i}{n}} \AtN{u^{\ast}_{i}}{n-1/2}$.

  \item Perform the implicit geodesic velocity update at fixed $\AtN{x^i}{n}$:
  \item[] $\AtN{u^{\ast}_{i}}{n-1/2} \xRightarrow[\Delta t]{\AtN{x^i}{n}} \AtN{u^{\ast \ast}_{i}}{n-1/2}$.

  \item Transform the updated velocity back to the tetrad basis at $\AtN{x^i}{n}$:
  \item[] $\AtN{u^{\ast \ast}_i}{n-1/2} \xRightarrow[]{\AtN{x^i}{n}} \AtN{u^{\ast \ast}_{\hat i}}{n-1/2}$.

  \item Perform the second EM push (in tetrad basis) for \(\Delta t/2\):
  \item[] $\AtN{u^{\ast \ast}_{\hat i}}{n-1/2} \xRightarrow[\Delta t/2]{D_p^{\hat i}, B_p^{\hat i}} \AtN{u_{\hat i}}{n+1/2}$.

  \item Transform the updated velocity to covariant basis at $\AtN{x^i}{n}$; this completes the velocity half of the leapfrog update:
  \item[] $\AtN{u_{\hat i}}{n+1/2} \xRightarrow[]{\AtN{x^i}{n}} \AtN{u_{i}}{n+1/2}$.

  \item Perform the implicit geodesic coordinate update (full \(\Delta t\)):
  \item[] $\AtN{x^i}{n} \xRightarrow[\Delta t]{ \AtN{u_{i}}{n+1/2}} \AtN{x^i}{n+1}$.

  \item Update the out-of-plane coordinate using \eqref{eq:discretized-geodesic-x-phi}:
  \item[] $\AtN{\phi}{n} \xRightarrow[\Delta t]{\AtN{x^{i}}{n+1/2}, \AtN{u_{i}}{n+1/2}} \AtN{\phi}{n+1}$.

\end{enumerate}
Steps 4--6 and 8--10 are part of the \texttt{EMpush} routine within the particle pusher kernel of the GR module. The geodesic velocity and coordinate updates are performed by the \texttt{GeodesicMomentumPush} and \texttt{GeodesicCoordinatePush} routines, respectively. For the massless particles, the \texttt{EMpush} routine is skipped.

\subsection{Electric currents}
\label{sec:currents}

We derive the charge continuity equation by taking the divergence of the covariant Amp\`ere's law and using Poisson's law:
\begin{align}
  \begin{aligned}
    \frac{\partial}{\partial t} \partial_i \left( \sqrt{h}\, D^i \right) & = -4\pi\, \partial_i \left( \sqrt{h}\, J^i \right), \\
    \frac{1}{\sqrt{h}}\, \partial_i \left( \sqrt{h}\, D^i \right)        & = 4\pi\, \rho.
  \end{aligned}
\end{align}
Here, $\rho$ denotes the charge density. Therefore, the charge conservation takes the same form as in the special relativity (see paper I):
\begin{equation}
  \label{eq:charge-conservation}
  \frac{\partial}{\partial t} \left( \sqrt{h}\, \rho \right) + \partial_i \left( \sqrt{h}\, J^i \right) = 0.
\end{equation}

As discussed in paper~I, we employ a coordinate-conformal particle shape function $S(x^i - x_p^i)$ and deposit conformal currents $\mathcal{J}^i \equiv \sqrt{h}\, J^i$ on the grid. Here, $x^i$ denotes the grid coordinate, and $x_p^i$ is the position of each macroparticle. The total charge density contributed by all particles at a given location is
\begin{equation}
  \rho = \sum_s q_s \sum_p \frac{1}{\sqrt{h}}\, S(x^i - x_p^i),
\end{equation}
where $\sum_p$ runs over all particles of species $s$, and $\sum_s$ is the sum over all species. Then the time-discretized form of the current deposition is given by
\begin{equation}
  \label{eq:charge-conservation-disc}
  \begin{split}
    \partial_i \AtN{\mathcal{J}^i}{n+1/2} = &
    - \sum_s q_s \cdot                               \\
                                            & \sum_p
    \frac{ S(x^i - \AtN{x_p^i}{n+1}) - S(x^i - \AtN{x_p^i}{n}) }{ \Delta t }.
  \end{split}
\end{equation}
Thus, the current deposition and filtering routines are identical to those used in the special-relativistic curvilinear module.

From Eq.~\eqref{eq:charge-conservation-disc}, it follows that only the initial and final particle positions, $\AtN{x_p^i}{n}$ and $\AtN{x_p^i}{n+1}$, are required for the current deposition. In explicit PIC schemes, the final position is known, and the initial one can typically be inferred as
\begin{equation}
  \AtN{x_p^i}{n} = \AtN{x_p^i}{n+1} - \Delta t\, \AtN{u_i}{n+1/2}.
\end{equation}

\noindent However, in \entity, this approach cannot be used, and the initial positions must instead be stored explicitly.

For the out-of-plane current density, on the other hand, we can compute its value directly, using the following formula
\begin{equation}
  J^3=q_s\frac{d x^3}{d t} = q_s\frac{\alpha}{\gamma}\, u^3,
\end{equation}
which is identical to the special-relativistic expression, except for the inclusion of the lapse function $\alpha \neq 1$.

\subsection{Field solver}
In $3\!+\!1$ formulation, the electromagnetic field evolution is written in covariant form as
\begin{equation}
  \begin{aligned}
    \frac{\partial D^i}{\partial t} & = e^{ijk}\partial_j H_k - 4\pi J^i, \\
    \frac{\partial B^i}{\partial t} & = -e^{ijk}\partial_j E_k,
  \end{aligned}
  \label{eq:fields}
\end{equation}
with the auxiliary fields determined as:
\begin{equation}
  \begin{aligned}
    H_i & = \alpha\,h_{ij}B^j - e_{ijk}\beta^j D^k, \\
    E_i & = \alpha\,h_{ij}D^j + e_{ijk}\beta^j B^k,
  \end{aligned}
  \label{eq:aux}
\end{equation}
where $e_{ijk}=\sqrt{h}\,\varepsilon_{ijk}$ and $\varepsilon_{ijk}$ is the Levi--Civita symbol.

Compared to special-relativistic PIC, GR evolution requires storing additional time-shifted field states for interpolation and time-centering. In \entity, the arrays denoted as \texttt{em0} (see paper I for more details on how the fields are stored) correspond to $\AtN{B^i}{n-3/2}$ and $\AtN{D^i}{n-1}$, while the auxiliary arrays, \texttt{aux}, store $\AtN{H_i}{n-1/2}$ and $\AtN{E_i}{n}$. Currents are defined at $\AtN{J^i}{n+1/2}$ (computed from particle positions $\AtN{x^i_p}{n}$ and $\AtN{x^i_p}{n+1}$) and we also retain the previous-step currents, \texttt{cur0}, $\AtN{J^i}{n-1/2}$. Because $H_i$ and $E_i$ depend both on $B^i$ and $D^i$, the magnetic and electric field updates are coupled and must be known simultaneously to calculate the auxiliary fields, complicating the canonical leapfrog update \citep{Parfrey2019PhRvL}.

\subsubsection{Spatial discretization}
\label{sec:spatial-disc}


\begin{figure*}
  \centering
  \resizebox{\textwidth}{!}{
    \begin{tikzpicture}[]
      \tikzstyle{every node}=[font=\normalsize]

      \definecolor{bfieldcolor}{HTML}{0474BA}
      \definecolor{efieldcolor}{HTML}{F17720}
      \definecolor{dimfieldcolor}{HTML}{7f7f7f}
      \definecolor{ghostcolor}{HTML}{e1e1e1}

      \def\lw{1pt}
      \def\arrowLw{1.5pt}
      \def\arrowGap{0.05}
      \def\circGap{0.1}
      \def\circR{0.1}
      \def\size{2}

      \newcommand{\veccircle}[6]{%
        \pgfmathsetmacro{\X}{(#1 + #3)*\size}
        \pgfmathsetmacro{\Y}{(#2 + #4)*\size}
        \pgfmathsetmacro{\R}{(#5)*\size}

        \draw[color=#6, line width=\lw]
        (\X, \Y) circle[radius=\R];

        \draw[color=#6, line width=\lw/2]
        (\X-0.7*\R, \Y-0.7*\R) -- (\X+0.7*\R, \Y+0.7*\R);

        \draw[color=#6, line width=\lw/2]
        (\X-0.7*\R, \Y+0.7*\R) -- (\X+0.7*\R, \Y-0.7*\R);
      }

      \newcommand{\vecXE}[7]{
        \pgfmathsetmacro{\Xc}{(#1 + #3)*\size}
        \pgfmathsetmacro{\Yc}{(#2 + #4)*\size}
        \pgfmathsetmacro{\Len}{(#5)*\size}
        \pgfmathsetmacro{\Xs}{\Xc - \Len/2}
        \pgfmathsetmacro{\Xe}{\Xc + \Len/2}
        \draw [color=#7, line width=\arrowLw, #6] (\Xs, \Yc) -- (\Xe, \Yc);
      }

      \newcommand{\vecXF}[7]{
        \pgfmathsetmacro{\Xs}{(#1 + #3)*\size}
        \pgfmathsetmacro{\Ys}{(#2 + #4)*\size}
        \pgfmathsetmacro{\Len}{(#5)*\size}
        \pgfmathsetmacro{\Xe}{\Xs + \Len/1.5}
        \draw [color=#7, line width=\arrowLw, #6] (\Xs, \Ys) -- (\Xe, \Ys);
      }

      \newcommand{\vecYE}[7]{
        \pgfmathsetmacro{\Xc}{(#1 + #3)*\size}
        \pgfmathsetmacro{\Yc}{(#2 + #4)*\size}
        \pgfmathsetmacro{\Len}{(#5)*\size}
        \pgfmathsetmacro{\Ys}{\Yc - \Len/2}
        \pgfmathsetmacro{\Ye}{\Yc + \Len/2}
        \draw [color=#7, line width=\arrowLw, #6] (\Xc, \Ys) -- (\Xc, \Ye);
      }

      \newcommand{\vecYF}[7]{
        \pgfmathsetmacro{\Xs}{(#1 + #3)*\size}
        \pgfmathsetmacro{\Ys}{(#2 + #4)*\size}
        \pgfmathsetmacro{\Len}{(#5)*\size}
        \pgfmathsetmacro{\Ye}{\Ys + \Len/1.5}
        \draw [color=#7, line width=\arrowLw, #6] (\Xs, \Ys) -- (\Xs, \Ye);
      }

      \newcommand{\drawgrid}[4]{%
        \pgfmathsetmacro{\ncols}{#1}
        \pgfmathsetmacro{\nrows}{#2}
        \pgfmathsetmacro{\cornerX}{#3}
        \pgfmathsetmacro{\cornerY}{#4}
        \foreach \i in {0,...,\ncols} {
            \draw (\cornerX*\size + \i*\size, \cornerY*\size) -- (\cornerX*\size + \i*\size, \cornerY*\size + \nrows*\size);
          }
        \foreach \j in {0,...,\nrows} {
            \draw (\cornerX*\size, \cornerY*\size + \j*\size) -- (\cornerX*\size + \ncols*\size, \cornerY*\size + \j*\size);
          }
      }


      \fill[fill=ghostcolor] (0, 0) rectangle (\size, 2*\size);
      \fill[fill=ghostcolor] (1.25*\size, 0) rectangle (2.25*\size, 2*\size);
      \fill[fill=ghostcolor] (0, 4.5*\size) rectangle (\size, 6.5*\size);
      \fill[fill=ghostcolor] (1.25*\size, 4.5*\size) rectangle (2.25*\size, 6.5*\size);
      \fill[fill=ghostcolor] (3*\size, 0) rectangle (4*\size, 3*\size);
      \fill[fill=ghostcolor] (3*\size, 3.5*\size) rectangle (4*\size, 6.5*\size);
      \fill[fill=ghostcolor] (6.25*\size, 0) rectangle (7.25*\size, 3*\size);
      \fill[fill=ghostcolor] (6.25*\size, 3.5*\size) rectangle (7.25*\size, 6.5*\size);

      \node [] at (4.5*\size, 6.75*\size) {Faraday: $B^i$};
      \node [] at (7.75*\size, 6.75*\size) {Amp\'ere: $D^i$};
      \node [] at (4.5*\size, 3.25*\size) {$H_i$};
      \node [] at (7.75*\size, 3.25*\size) {$E_i$};

      {
      \def\xOffset{0}
      \def\yOffset{2}

      \node [left] at (\xOffset*\size, \yOffset*\size + 0.15*\size) {$\theta=0$};
      \node [left] at (\xOffset*\size, \yOffset*\size - 0.15*\size) {$\mathtt{i_2}=0$};

      \drawgrid{1}{3}{\xOffset}{\yOffset-2}
      \veccircle{\xOffset}{\yOffset}{0.5}{0.5}{\circR}{bfieldcolor}
      \veccircle{\xOffset}{\yOffset}{0.5}{-0.5}{\circR}{bfieldcolor}
      \vecXF{\xOffset}{\yOffset}{0}{0.5}{0.5}{->}{bfieldcolor}
      \vecXF{\xOffset}{\yOffset}{0}{-0.5}{0.5}{->}{bfieldcolor}
      \vecYF{\xOffset}{\yOffset}{0.5}{-1}{0.5}{->}{bfieldcolor}
      \vecYF{\xOffset}{\yOffset}{0.5}{1}{-0.5}{->}{bfieldcolor}
      \vecYE{\xOffset}{\yOffset}{0.5}{0}{0.25}{-}{bfieldcolor}

      \node [above] at (\xOffset*\size + 0.5*\size, \yOffset*\size + 1.1*\size) {$\vdots$};
      }

      {
      \def\xOffset{1.25}
      \def\yOffset{2}
      \drawgrid{1}{3}{\xOffset}{\yOffset-2}
      \fill[color=efieldcolor] (\xOffset*\size, \yOffset*\size) circle[radius=\circR];
      \veccircle{\xOffset}{\yOffset}{0}{-1}{\circR}{efieldcolor}
      \veccircle{\xOffset}{\yOffset}{0}{1}{\circR}{efieldcolor}
      \vecXE{\xOffset}{\yOffset}{0.5}{-1}{0.5}{->}{efieldcolor}
      \vecXE{\xOffset}{\yOffset}{0.5}{1}{0.5}{->}{efieldcolor}
      \vecYE{\xOffset}{\yOffset}{0}{0.5}{0.5}{->}{efieldcolor}
      \vecYE{\xOffset}{\yOffset}{0}{-0.5}{0.5}{<-}{efieldcolor}

      \node [above] at (\xOffset*\size + 0.5*\size, \yOffset*\size + 1.1*\size) {$\vdots$};
      }

      {
      \def\xOffset{4}
      \def\yOffset{1}
      \drawgrid{3}{3}{\xOffset-1}{\yOffset-1}
      \node [right] at (\xOffset*\size - \size, \yOffset*\size - 0.8*\size) {$\mathtt{i_1}=-1$};
      \node [right] at (\xOffset*\size, \yOffset*\size - 0.8*\size) {$\mathtt{i_1}=0$};
      \node [right] at (\xOffset*\size + \size, \yOffset*\size - 0.8*\size) {$\mathtt{i_1}=1$};

      \node at (\xOffset*\size + 0.5*\size, \yOffset*\size + 1.75*\size)
      {$\textcolor{dimfieldcolor}{H_i},~\textcolor{bfieldcolor}{B^i},~\textcolor{efieldcolor}{D^i}$};

      \veccircle{\xOffset}{\yOffset}{0}{0}{\circR}{efieldcolor}
      \veccircle{\xOffset}{\yOffset}{1}{0}{\circR}{efieldcolor}
      \veccircle{\xOffset}{\yOffset}{-0.5}{0.5}{\circR}{bfieldcolor}
      \veccircle{\xOffset}{\yOffset}{0.5+\circGap}{0.5}{\circR}{bfieldcolor}
      \veccircle{\xOffset}{\yOffset}{0.5-\circGap}{0.5}{\circR}{dimfieldcolor}
      \vecYE{\xOffset}{\yOffset}{0}{0.5}{0.5}{->}{efieldcolor}
      \vecYE{\xOffset}{\yOffset}{1}{0.5}{0.5}{->}{efieldcolor}
      \vecYF{\xOffset}{\yOffset}{0.5+\arrowGap}{0}{0.5}{->}{bfieldcolor}
      \vecYF{\xOffset}{\yOffset}{0.5-\arrowGap}{0}{0.5}{->}{dimfieldcolor}
      \vecXF{\xOffset}{\yOffset}{0}{0.5+\arrowGap}{0.5}{->}{bfieldcolor}
      \vecXF{\xOffset}{\yOffset}{0}{0.5-\arrowGap}{0.5}{->}{dimfieldcolor}
      \vecXF{\xOffset}{\yOffset}{1}{0.5}{0.5}{->}{bfieldcolor}
      }

      {
      \def\xOffset{7.25}
      \def\yOffset{1}
      \drawgrid{3}{3}{\xOffset-1}{\yOffset-1}
      \node [right] at (\xOffset*\size - \size, \yOffset*\size - 0.8*\size) {$\mathtt{i_1}=-1$};
      \node [right] at (\xOffset*\size, \yOffset*\size - 0.8*\size) {$\mathtt{i_1}=0$};
      \node [right] at (\xOffset*\size + \size, \yOffset*\size - 0.8*\size) {$\mathtt{i_1}=1$};

      \node at (\xOffset*\size + 0.5*\size, \yOffset*\size + 1.75*\size)
      {$\textcolor{dimfieldcolor}{E_i},~\textcolor{efieldcolor}{D^i},~\textcolor{bfieldcolor}{B^i}$};

      \veccircle{\xOffset}{\yOffset}{\circGap}{0}{\circR}{efieldcolor}
      \veccircle{\xOffset}{\yOffset}{-\circGap}{0}{\circR}{dimfieldcolor}
      \veccircle{\xOffset}{\yOffset}{1}{0}{\circR}{efieldcolor}
      \veccircle{\xOffset}{\yOffset}{0.5}{0.5}{\circR}{bfieldcolor}
      \veccircle{\xOffset}{\yOffset}{-0.5}{0.5}{\circR}{bfieldcolor}
      \vecYE{\xOffset}{\yOffset}{\arrowGap}{0.5}{0.5}{->}{efieldcolor}
      \vecYE{\xOffset}{\yOffset}{-\arrowGap}{0.5}{0.5}{->}{dimfieldcolor}
      \vecXE{\xOffset}{\yOffset}{-0.5}{0}{0.5}{->}{efieldcolor}
      \vecXE{\xOffset}{\yOffset}{0.5}{\arrowGap}{0.5}{->}{efieldcolor}
      \vecXE{\xOffset}{\yOffset}{0.5}{-\arrowGap}{0.5}{->}{dimfieldcolor}
      \vecYF{\xOffset}{\yOffset}{-0.5}{0}{0.5}{->}{bfieldcolor}
      \vecYF{\xOffset}{\yOffset}{0.5}{0}{0.5}{->}{bfieldcolor}
      }

      {
      \def\xOffset{0}
      \def\yOffset{4.5}

      \node [color=bfieldcolor] at (\xOffset*\size+0.5*\size, \yOffset*\size + 2.25*\size) {$B^i$};

      \node [left] at (\xOffset*\size, \yOffset*\size + 0.15*\size) {$\theta=\pi$};
      \node [left] at (\xOffset*\size, \yOffset*\size - 0.15*\size) {$\mathtt{i_2}=\mathtt{N_\theta}$};

      \drawgrid{1}{3}{\xOffset}{\yOffset-1}
      \veccircle{\xOffset}{\yOffset}{0.5}{0.5}{\circR}{bfieldcolor}
      \veccircle{\xOffset}{\yOffset}{0.5}{-0.5}{\circR}{bfieldcolor}
      \vecYE{\xOffset}{\yOffset}{0.5}{0}{0.25}{-}{bfieldcolor}
      \vecXF{\xOffset}{\yOffset}{0}{0.5}{0.5}{->}{bfieldcolor}
      \vecXF{\xOffset}{\yOffset}{0}{-0.5}{0.5}{->}{bfieldcolor}
      \vecYF{\xOffset}{\yOffset}{0.5}{1}{-0.5}{->}{bfieldcolor}
      \vecYF{\xOffset}{\yOffset}{0.5}{-1}{0.5}{->}{bfieldcolor}
      }

      {
      \def\xOffset{1.25}
      \def\yOffset{4.5}
      \drawgrid{1}{3}{\xOffset}{\yOffset-1}

      \node [color=efieldcolor] at (\xOffset*\size+0.5*\size, \yOffset*\size + 2.25*\size) {$D^i$};

      \fill[color=efieldcolor] (\xOffset*\size, \yOffset*\size) circle[radius=\circR];
      \veccircle{\xOffset}{\yOffset}{0}{1}{\circR}{efieldcolor}
      \veccircle{\xOffset}{\yOffset}{0}{-1}{\circR}{efieldcolor}
      \vecXE{\xOffset}{\yOffset}{0.5}{-1}{0.5}{->}{efieldcolor}
      \vecXE{\xOffset}{\yOffset}{0.5}{1}{0.5}{->}{efieldcolor}
      \vecYE{\xOffset}{\yOffset}{0}{0.5}{0.5}{<-}{efieldcolor}
      \vecYE{\xOffset}{\yOffset}{0}{-0.5}{0.5}{->}{efieldcolor}
      }

      {
      \def\xOffset{4}
      \def\yOffset{4.5}
      \drawgrid{3}{3}{\xOffset-1}{\yOffset-1}

      \node at (\xOffset*\size + 0.5*\size, \yOffset*\size + 1.75*\size)
      {$\textcolor{dimfieldcolor}{B^i},~\textcolor{efieldcolor}{D^i}$};

      \veccircle{\xOffset}{\yOffset}{0}{0}{\circR}{efieldcolor}
      \veccircle{\xOffset}{\yOffset}{0}{1}{\circR}{efieldcolor}
      \veccircle{\xOffset}{\yOffset}{1}{0}{\circR}{efieldcolor}
      \veccircle{\xOffset}{\yOffset}{0.5}{0.5}{\circR}{dimfieldcolor}
      \vecYE{\xOffset}{\yOffset}{0}{0.5}{0.5}{->}{efieldcolor}
      \vecYE{\xOffset}{\yOffset}{1}{0.5}{0.5}{->}{efieldcolor}
      \vecXE{\xOffset}{\yOffset}{0.5}{1}{0.5}{->}{efieldcolor}
      \vecXE{\xOffset}{\yOffset}{0.5}{0}{0.5}{->}{efieldcolor}
      \vecXF{\xOffset}{\yOffset}{0}{0.5}{0.5}{->}{dimfieldcolor}
      \vecYF{\xOffset}{\yOffset}{0.5}{0}{0.5}{->}{dimfieldcolor}
      }

      {
      \def\xOffset{7.25}
      \def\yOffset{4.5}
      \drawgrid{3}{3}{\xOffset-1}{\yOffset-1}

      \node at (\xOffset*\size + 0.5*\size, \yOffset*\size + 1.75*\size)
      {$\textcolor{dimfieldcolor}{D^i},~\textcolor{bfieldcolor}{B^i}$};

      \veccircle{\xOffset}{\yOffset}{0.5}{0.5}{\circR}{bfieldcolor}
      \veccircle{\xOffset}{\yOffset}{-0.5}{0.5}{\circR}{bfieldcolor}
      \veccircle{\xOffset}{\yOffset}{0.5}{-0.5}{\circR}{bfieldcolor}
      \veccircle{\xOffset}{\yOffset}{0}{0}{\circR}{dimfieldcolor}
      \vecXE{\xOffset}{\yOffset}{0.5}{0}{0.5}{->}{dimfieldcolor}
      \vecYE{\xOffset}{\yOffset}{0}{0.5}{0.5}{->}{dimfieldcolor}
      \vecXF{\xOffset}{\yOffset}{0}{0.5}{0.5}{->}{bfieldcolor}
      \vecYF{\xOffset}{\yOffset}{0.5}{0}{0.5}{->}{bfieldcolor}
      \vecXF{\xOffset}{\yOffset}{0}{-0.5}{0.5}{->}{bfieldcolor}
      \vecYF{\xOffset}{\yOffset}{-0.5}{0}{0.5}{->}{bfieldcolor}
      }

    \end{tikzpicture}
  }
  \caption{\textit{Left:} locations of $B^i$ and $D^i$ components for which boundary conditions on both of the axes are defined. The solid (no arrow) lines for $B^2_{(\mathtt{i+1/2},0)}$, $B^2_{(\mathtt{i+1/2},\mathtt{N_\theta})}$, as well as solid circles for $D^3_{(i,0)}$, and $D^3_{(i,\mathtt{N_\theta})}$ indicate that these components are set to zero by symmetry. The other components are either reflected exactly ($r$-components) or reflected with the opposite sign ($\theta$-components). \textit{Right:} Location of the components of electromagnetic fields used for each of the four main routines. The deduced components for the cell $(\mathtt{i_{1}=0}, \mathtt{i_2})$ are shown with gray. All the magnetic and electric field components used in the respective routines are shown in blue and orange, respectively.}
  \label{fig:spat-disc-cartgrid}
\end{figure*}

The spatial discretization of $B^i$, $D^i$, and $J^i$ is done according to the Yee scheme \citep{Yee_1966}, with auxiliary fields $H_i$ and $E_i$ discretized in the same manner as the primary fields. In the Yee mesh, the electromagnetic fields $B^i$ and $D^i$ are defined at different spatial locations, requiring interpolation to a common location for the computation of $H_i$ and $E_i$, in addition to interpolation in time. An additional complication arises from the presence of a non-diagonal spatial metric, which requires interpolation of different components of the same field. The computational grid is uniform and equally spaced in the general code coordinates. The space-time geometry is specified by the metric coefficients.

In this section, we present the discretized forms of Faraday and Amp\`ere routines, given by Eqs.~\eqref{eq:fields}, along with the computation of the auxiliary fields in Eq.~\eqref{eq:aux}; see the right panel of Figure~\ref{fig:spat-disc-cartgrid}. Each of the Faraday and Amp\`ere substeps in the cycle described in Section 3.2 is carried out in the same manner. The discretized Faraday closely resembles the curvilinear SRPIC version described in paper I, but here includes the additional contributions due to the non-diagonal metric.

In the equations below, we use the following conventions for brevity:
\begin{equation}
  \begin{aligned}
     & \Delta_{\mathtt{n}}^{\mathtt{n+1}}\left[{}^{(\mathtt{*})}F\right] \equiv \frac{{}^{(\mathtt{n+1})}F-{}^{(\mathtt{n})}F}{\Delta t},   \\
     & \Delta_{\mathtt{i}}^{\mathtt{i+1}}\left[F_{(\mathtt{*,j})}\right] \equiv \frac{F_{(\mathtt{i+1,j})}-F_{(\mathtt{i,j})}}{\Delta x^1}, \\
     & \Delta_{\mathtt{j}}^{\mathtt{j+1}}\left[F_{(\mathtt{i,*})}\right] \equiv \frac{F_{(\mathtt{i,j+1})}-F_{(\mathtt{i,j})}}{\Delta x^2}. \\
  \end{aligned}
\end{equation}
The following describes the main Faraday update (see step 6 in Sec.~\ref{sec:pic_loop}):
\begin{widetext}
  \label{eq:discretized-faraday}
  \begin{align}
    \begin{aligned}
       & \Delta_{\mathtt{n-1/2}}^{\mathtt{n+1/2}}
      \left[\AtNIJ{B^1}{*}{i}{\jPhalf}\right]=
      -\frac{
        \Delta_{\mathtt{j}}^{\mathtt{j+1}}\left[
          \CovAtNIJ{E}{3}{n}{i}{*}
          \right]
      }{\sqrt{h_{(\mathtt{i,\jPhalf})}}}
      ,                                           \\
       & \Delta_{\mathtt{n-1/2}}^{\mathtt{n+1/2}}
      \left[\AtNIJ{B^2}{*}{\iPhalf}{j}\right]=
      \frac{
        \Delta_{\mathtt{i}}^{\mathtt{i+1}}\left[
          \CovAtNIJ{E}{3}{n}{*}{j}
          \right]
      }{\sqrt{h_{(\mathtt{\iPhalf,j})}}}
      ,                                           \\
       & \Delta_{\mathtt{n-1/2}}^{\mathtt{n+1/2}}
      \left[\AtNIJ{B^3}{*}{\iPhalf}{\jPhalf}\right]=
      -\frac{
        \Delta_{\mathtt{i}}^{\mathtt{i+1}}\left[
          \CovAtNIJ{E}{2}{n}{*}{\jPhalf}
          \right]-
        \Delta_{\mathtt{j}}^{\mathtt{j+1}}\left[
          \CovAtNIJ{E}{1}{n}{\iPhalf}{*}
          \right]
      }
      {\sqrt{h_{(\mathtt{\iPhalf,\jPhalf})}}}
      .
    \end{aligned}
  \end{align}
\end{widetext}

The update due to the Amp\`ere's law (see step 7 in Sec.~\ref{sec:pic_loop}), where $\mathcal{J}^i$ are the deposited conformal currents, is:
\begin{widetext}
  \label{eq:discretized-ampere}
  \begin{align}
    \begin{aligned}
       & \Delta_{\mathtt{n}}^{\mathtt{n+1}}
      \left[\AtNIJ{D^1}{*}{\iPhalf}{j}\right]=
      \frac{
        \Delta_{\mathtt{j-1/2}}^{\mathtt{j+1/2}}\left[
          \CovAtNIJ{H}{3}{n+1/2}{\iPhalf}{*}
          \right]
        - 4 \pi \AtNIJ{\mathcal{J}^1}{n+1/2}{\iPhalf}{j}
      }{\sqrt{h_{(\mathtt{\iPhalf,j})}}}
      ,                                     \\
       & \Delta_{\mathtt{n}}^{\mathtt{n+1}}
      \left[\AtNIJ{D^2}{*}{i}{\jPhalf}\right]=
      \frac{
        -\Delta_{\mathtt{i-1/2}}^{\mathtt{i+1/2}}\left[
          \CovAtNIJ{H}{3}{n+1/2}{*}{\jPhalf}
          \right]
        - 4 \pi \AtNIJ{\mathcal{J}^2}{n+1/2}{i}{\jPhalf}
      }{\sqrt{h_{(\mathtt{i,\jPhalf)}}}}
      ,                                     \\
       & \Delta_{\mathtt{n}}^{\mathtt{n+1}}
      \left[\AtNIJ{D^3}{*}{i}{j}\right]=
      \frac{\left\{
        \Delta_{\mathtt{i-1/2}}^{\mathtt{i+1/2}}\left[
          \CovAtNIJ{H}{2}{n+1/2}{*}{j}
          \right]-
        \Delta_{\mathtt{j-1/2}}^{\mathtt{j+1/2}}\left[
          \CovAtNIJ{H}{1}{n+1/2}{i}{*}
          \right]\right\}
        - 4 \pi \AtNIJ{\mathcal{J}^3}{n+1/2}{i}{j}
      }{\sqrt{h_{(\mathtt{i,j})}}}.
    \end{aligned}
  \end{align}
\end{widetext}

The spatial discretization of the auxiliary electric field $E_i$ and magnetic field $H_i$ is given by:
\begin{widetext}
  \begin{equation}
    \label{eq:discretized-auxE}
    \begin{aligned}
       & \begin{cases}
           \CovAtNIJ{E}{1}{n}{\iPhalf}{j}=
           \AtIJ{\alpha}{\iPhalf}{j} \left[
           \CovAtIJ{h}{11}{\iPhalf}{j} \AtNIJ{D^1}{n}{\iPhalf}{j}
           + {\AtNIJ{\langle h_{13}D^3 \rangle }{n}{\iPhalf}{j}}
           \right],                                                               \\
           \CovAtNIJ{E}{2}{n}{i}{\jPhalf}=
           \AtIJ{\alpha}{i}{\jPhalf} \left[
             \CovAtIJ{h}{22}{i}{\jPhalf} \AtNIJ{D^2}{n}{i}{\jPhalf}
           \right] - {\AtNIJ{\langle \sqrt{h}\beta_1B^3 \rangle}{n}{i}{\jPhalf}}, \\
           \CovAtNIJ{E}{3}{n}{i}{j}=
           \AtIJ{\alpha}{i}{j} \left[
           \CovAtIJ{h}{33}{i}{j} \AtNIJ{D^3}{n}{i}{j}
           + {\AtNIJ{\langle h_{13}D^1 \rangle}{n}{i}{j}}
           \right] + {\AtNIJ{\langle \sqrt{h}\beta^1 B^2 \rangle}{n}{i}{j}},      \\
         \end{cases} \\
       & \begin{cases}
           \CovAtNIJ{H}{1}{n}{i}{\jPhalf}=
           \AtIJ{\alpha}{i}{\jPhalf} \left[
           \CovAtIJ{h}{11}{i}{\jPhalf} \AtNIJ{B^1}{n}{i}{\jPhalf}
           + {\AtNIJ{\langle h_{13}B^3 \rangle}{n}{i}{\jPhalf}}
           \right],                                                               \\
           \CovAtNIJ{H}{2}{n}{\iPhalf}{j}=
           \AtIJ{\alpha}{\iPhalf}{j} \left[
             \CovAtIJ{h}{22}{\iPhalf}{j} \AtNIJ{B^2}{n}{\iPhalf}{j}
           \right] + \AtNIJ{\langle \sqrt{h} \beta^1 D^3 \rangle}{n}{\iPhalf}{j}, \\
           \begin{aligned}
          \CovAtNIJ{H}{3}{n}{\iPhalf}{\jPhalf}=
          \AtIJ{\alpha}{\iPhalf}{\jPhalf} \left[
            \CovAtIJ{h}{33}{\iPhalf}{\jPhalf} \AtNIJ{B^3}{n}{\iPhalf}{\jPhalf}
            + \AtNIJ{\langle h_{13} B^1 \rangle}{n}{\iPhalf}{\jPhalf}
          \right] \\ - \AtNIJ{\langle \sqrt{h} \beta^1 D^2 \rangle}{n}{\iPhalf}{\jPhalf}.
        \end{aligned}
         \end{cases}
    \end{aligned}
  \end{equation}
\end{widetext}

\noindent Here, the quantities enclosed in angular brackets $\langle \cdot \rangle$ are not defined exactly at the locations where they are required and therefore must be interpolated. We perform this interpolation using a volume-weighted averaging scheme. For any quantity $a$, the value at a grid point $({\mathtt i},{\mathtt j})$ is computed as:
\begin{equation}
  \AtIJ{V}{i}{j} \AtIJ{a}{i}{j} = \frac{1}{2} \left(\AtIJ{V}{i+1}{j} \AtIJ{a}{i+1}{j} + \AtIJ{V}{i-1}{j} \AtIJ{a}{i-1}{j} \right).
\end{equation}

\noindent Therefore, the value of $a$ at a radial location ${\mathtt i}$ is obtained from its neighboring values at ${\mathtt i-1}$ and ${\mathtt i+1}$, weighted by their corresponding cell volumes $\AtIJ{V}{i\pm1}{j}$. The volume of a computational cell is given by $\AtIJ{V}{i}{j} = \sqrt{\AtIJ{h}{i}{j}} dx_1 dx_2$, where $dx_1 = dx_2 = 1$ for our choice of a uniform grid spacing.

The explicit form for the interpolated quantities in \eqref{eq:discretized-auxE} are then
\begin{widetext}
  \begin{equation}
    \label{eq:discretized-auxH}
    \begin{aligned}
      {\AtNIJ{\langle h_{13}D^3 \rangle}{n}{\iPhalf}{j}}                & = \frac{\AtIJ{[\sqrt{\tilde{h}} h_{13}]}{i}{j}\AtNIJ{D^3}{n}{i}{j} + \AtIJ{[\sqrt{\tilde{h}} h_{13}]}{i+1}{j} \AtNIJ{D^3}{n}{i+1}{j}}{2\AtIJ{\sqrt{\tilde{h}}}{\iPhalf}{j}},                                                          \\
      {\AtNIJ{\langle h_{13}B^3 \rangle}{n}{i}{\jPhalf}}                & = \frac{\AtIJ{[\sqrt{\tilde{h}}h_{13}]}{\iMhalf}{\jPhalf} \AtNIJ{B^3}{n}{\iMhalf}{\jPhalf} + \AtIJ{[\sqrt{\tilde{h}}h_{13}]}{\iPhalf}{\jPhalf} \AtNIJ{B^3}{n}{\iPhalf}{\jPhalf}}{2\AtIJ{\sqrt{\tilde{h}}}{i}{\jPhalf}},               \\
      \AtNIJ{ \langle h_{13}D^1 \rangle}{n}{i}{j}                       & = \frac{\AtIJ{[\sqrt{\tilde{h}}h_{13}]}{\iMhalf}{j} \AtNIJ{D^1}{n}{\iMhalf}{j} + \AtIJ{[\sqrt{\tilde{h}}h_{13}]}{\iPhalf}{j} \AtNIJ{D^1}{n}{\iPhalf}{j}}{2 \AtIJ{\sqrt{\tilde{h}}}{i}{j}},                                            \\
      {\AtNIJ{ \langle \sqrt{h} \beta^1 B^2 \rangle}{n}{i}{j}}          & = \frac{\AtIJ{{[\sqrt{\tilde{h} h} \beta^1 ]}}{\iMhalf}{j} \AtNIJ{B^2}{n}{\iMhalf}{j} + \AtIJ{{[\sqrt{\tilde{h} h} \beta^1 ]}}{\iPhalf}{j} \AtNIJ{B^2}{n}{\iPhalf}{j}}{2\AtIJ{\sqrt{\tilde{h}}}{i}{j}},                               \\
      {\AtNIJ{\langle \sqrt{h} \beta^1 B^3 \rangle}{n}{i}{\jPhalf}}     & = \frac{\AtIJ{{[\sqrt{\tilde{h} h} \beta^1 ]}}{\iMhalf}{\jPhalf} \AtNIJ{B^3}{n}{\iMhalf}{\jPhalf} + \AtIJ{{[\sqrt{\tilde{h} h} \beta^1 ]}}{\iPhalf}{\jPhalf} \AtNIJ{B^3}{n}{\iPhalf}{\jPhalf}}{2\AtIJ{\sqrt{\tilde{h}}}{i}{\jPhalf}}, \\
      \AtNIJ{\langle \sqrt{h} \beta^1 D^3 \rangle}{n}{\iPhalf}{j}       & = \frac{\AtIJ{{[\sqrt{\tilde{h} h} \beta^1 ]}}{i}{j} \AtNIJ{D^3}{n}{i}{j} + \AtIJ{{[\sqrt{\tilde{h} h} \beta^1 ]}}{i+1}{j} \AtNIJ{D^3}{n}{i+1}{j}}{2\AtIJ{\sqrt{\tilde{h}}}{\iPhalf}{j}},                                             \\
      \AtNIJ{\langle h_{13} B^1 \rangle}{n}{\iPhalf}{\jPhalf}           & = \frac{\AtIJ{[\sqrt{\tilde{h}} h_{13}]}{i}{\jPhalf}\AtNIJ{B^1}{n}{i}{\jPhalf} + \AtIJ{[\sqrt{\tilde{h}} h_{13}]}{i+1}{\jPhalf} \AtNIJ{B^1}{n}{i+1}{\jPhalf}}{2\AtIJ{\sqrt{\tilde{h}}}{\iPhalf}{\jPhalf}},                            \\
      \AtNIJ{\langle \sqrt{h} \beta^1 D^2 \rangle}{n}{\iPhalf}{\jPhalf} & = \frac{\AtIJ{{[\sqrt{\tilde{h} h} \beta^1 ]}}{i}{\jPhalf} \AtNIJ{D^2}{n}{i}{\jPhalf} + \AtIJ{{[\sqrt{\tilde{h} h} \beta^1 ]}}{i+1}{\jPhalf} \AtNIJ{D^2}{n}{i+1}{\jPhalf}}{2\AtIJ{\sqrt{\tilde{h}}}{\iPhalf}{\jPhalf}}.
    \end{aligned}
  \end{equation}
\end{widetext}

\noindent Here, $\tilde{h}\equiv h / \sin^2{\theta}$, which is used as an auxiliary to avoid finite-precision errors; notice, that we are able to use this definition, since both the numerator and the denominator in all the relations from \eqref{eq:discretized-auxH} are defined at the same $\theta$-position, i.e., same index $\mathtt{j}$, and the potentially problematic $\sin$-s from the metric cancel out analytically.

\subsubsection{Special treatment of the polar axes}
Spherical coordinates contain a coordinate singularity at the axes ($\theta=0$ and $\theta=\pi$), where the metric determinant vanishes ($\sqrt{h}=0$). This raises an issue of how to evolve quantities defined exactly on the axes: in our discretization, these are the components $D^1$, $D^3$, and $B^2$. First of all, the $\theta$ and $\phi$ components of the field must vanish by symmetry -- these boundary conditions are described in detail in the next section. Consequently, we do not evolve $D^3$ and $B^2$ on the axes.

The component $D^1$ however, remains non-vanishing, and must therefore be updated in the Amp\'ere's law explicitly. As in the SR from paper I, we employ the integral form of the evolution equation for $D^1$, which involves evaluating $H_3$ that is symmetric with respect to the axis. The update then takes the form:
\begin{equation}
  \begin{split}
    \Delta_{\mathtt{n}}^{\mathtt{n+1}}
     & \left[\AtNIJ{D^1}{*}{\iPhalf}{j}\right]=                                                                           \\
     & \frac{\CovAtNIJ{H}{3}{n}{\iPhalf}{j\pm1/2} - 2\pi \AtNIJ{\mathcal{J}^1}{n+1/2}{\iPhalf}{j}}{A_{\mathtt{\iPhalf}}},
  \end{split}
\end{equation}
where we take $\CovAtNIJ{H}{3}{n}{\iPhalf}{j+1/2}$ in a cell next to the axis at $\theta=0$ ($\mathtt{i_2} = 0$), and $\CovAtNIJ{H}{3}{n}{\iPhalf}{j-1/2}$ in a cell next to the axis at $\theta=\pi$ (${\mathtt i_2} = \mathtt{N_\theta}$). As in SR, $A_{\mathtt{\iPhalf}}$ denotes the polar area with a half-cell-size radius near the axis: $A_\mathtt{\iPhalf}\equiv A(\mathtt{x^1}=\mathtt{\iPhalf})\equiv 2\pi \int_0^{1/2} \sqrt{h}~d\mathtt{x^2}$, the functional form of which is written explicitly in the Appendix~\ref{ap:metric}.

On exactly the axes, the metric component $h^{33}$ is singular, which can cause numerical issues in the geodesic pusher when particles approach the axis too close (due to finite numerical precision). In practice, this situation occurs rarely -- typically for only a few particles (out of billions) with nearly vanishing toroidal velocity. For such cases, at the beginning of each push, the particle physical coordinate $x^2$ is replaced with the value stored as $\texttt{SMALL\_ANGLE\_GR} \approx  10^{-5}$ (a compile-time definition) to avoid the singularity and ensure stable integration. The same regularization procedure is applied during the deposition of the out-of-plane current component, $\mathcal{J}^3$, where the same coordinate replacement is used for evaluating the particle's $u^3$ velocity component.

\subsubsection{Boundary conditions}
Boundary conditions are imposed on evolved electromagnetic field components, \texttt{em}, only. Under axisymmetry, fields in the ghost zones adjacent to the axis satisfy reflection symmetry (see the left panel of Figure~\ref{fig:spat-disc-cartgrid}). At the outer boundary, the fields are matched to their initial values using an absorbing layer (see \texttt{MATCH} boundaries described in paper I). Both the axial and the outer boundary conditions are identical to those in SR and are handled by the same routines.

The treatment for the particles, however, is different, since in SR their positions are updated in the global Cartesian basis, where the axis is not present. In GR, particles are ``reflected'' from the axes once their inferred coordinate lies outside the block: if $\texttt{i2}<0$ or $\texttt{i2} \ge \mathtt{N_\theta}$ (with $\texttt{i2}$ being the integer component of the particle position in code units). In that case, the cell index of the particle is reset to $\texttt{i2}=0$ or $\texttt{i2} = \mathtt{N_\theta-1}$, while the sub-cell displacement is updated as $\texttt{dx2} \to 1-\texttt{dx2}$, and the tangential velocity, $u_2$, is inverted, $u_2 \to -u_2$.

  {The computational domain extends inside the event horizon, where we impose constant boundary conditions for the electromagnetic fields. Specifically, all the field components in the layer of cells from $\texttt{i}_1 = -\texttt{N\_GHOSTS}$ to $\texttt{i}_1 = n_f$ are set equal to their corresponding values at $\texttt{i}_1 = n_f + 1$. Here, $n_f$ denotes the number of current filters applied. This treatment prevents filtered currents associated with particles removed within the horizon from coupling back into the physical domain. Consequently, the inner boundary of the computational domain must always be placed at least $n_f + 1$ cells inside the event horizon. Particles are allowed to cross the horizon freely and are removed from the computational domain once they pass through the layer of cells where constant boundary conditions for the fields are applied and enter the ghost zone.}



\subsection{PIC loop}\label{sec:pic_loop}

Each timestep proceeds according to the following scheme:
\begin{enumerate}
  \item Start at $\AtN{t}{n}$ with the following quantities stored:
        \begin{equation*}
          \begin{aligned}
             & \AtN{B^i}{n-1/2},~\AtN{D^i}{n};                         \\
             & \AtN{B^i}{n-3/2},~\AtN{D^i}{n-1};                       \\
             & \AtN{H_i}{n-1/2},~\AtN{E_i}{n};                         \\
             & \AtN{\mathcal{J}^i}{n-3/2},~\AtN{\mathcal{J}^i}{n-1/2}; \\
             & \AtN{u_i}{n-1/2},~\AtN{x^i}{n},~\AtN{x^i}{n-1},
          \end{aligned}
        \end{equation*}
        where $x^i$ and $u_i$ are the particle's coordinates and velocities, respectively.

  \item As in SR, we need to align $B^i$ and $D^i$ in time for the subsequent particle push, as both fields are part of the Lorentz force. This is done via several steps:
        \begin{itemize}

          \item Interpolate $B^i$ and $D^i$ to an intermediate time:
                \begin{equation}
                  \begin{aligned}
                     & \AtN{B^i}{n-1} = \frac{\AtN{B^i}{n-3/2} + \AtN{B^i}{n-1/2}}{2}; \\
                     & \AtN{D^i}{n-1/2} = \frac{\AtN{D^i}{n-1} + \AtN{D^i}{n}}{2}.
                  \end{aligned}
                \end{equation}

          \item Compute the auxiliary field $E_i$ at $\AtN{t}{n-1/2}$:
                \begin{equation}
                  \begin{aligned}
                    \{ \AtN{B^i}{n-1/2}, \AtN{D^i}{n-1/2} \} \rightarrow \AtN{E_i}{n-1/2}.
                  \end{aligned}
                \end{equation}

          \item Perform an auxiliary Faraday substep to update the interpolated magnetic field $\AtN{B^i}{n-1}$:
                \begin{equation}
                  \begin{aligned}
                    \AtN{B^i}{n-1}\xRightarrow[\Delta t]{\AtN{E^i}{n-1/2}} \AtN{B^i}{n}.
                  \end{aligned}
                \end{equation}

        \end{itemize}

  \item Using $\AtN{B^i}{n}$ and $\AtN{D^i}{n}$, we can now push the particles as outlined in Sec.~\ref{sec:particle-push}:
        \begin{equation}
          \begin{aligned}
            \AtN{u_i}{n-1/2} & \rightarrow \AtN{u_i}{n+1/2}; \\
            \AtN{x^i}{n}     & \rightarrow \AtN{x^i}{n+1}.
          \end{aligned}
        \end{equation}

  \item Using the initial and final locations of particles, the conformal current deposition is performed:
        \begin{equation}
          \{\AtN{x^i}{n}, ~\AtN{x^i}{n+1} \} \rightarrow \AtN{\mathcal{J}^i}{n+1/2}.
        \end{equation}

  \item We then compute the midpoint values for the electric currents:
        \begin{equation}
          \AtN{\mathcal{J}^i}{n} = \frac{\AtN{\mathcal{J}^i}{n-1/2} + \AtN{\mathcal{J}^i}{n+1/2}}{2}.
        \end{equation}

  \item Perform the full Faraday update for the $B^i$:
        \begin{itemize}
          \item
                Compute auxiliary fields at $\AtN{t}{n}$:
                \begin{equation}
                  \begin{aligned}
                    \{ \AtN{B^i}{n}, \AtN{D^i}{n} \} & \rightarrow \AtN{E_i}{n}; \\
                    \{ \AtN{B^i}{n}, \AtN{D^i}{n} \} & \rightarrow \AtN{H_i}{n}.
                  \end{aligned}
                \end{equation}

          \item Perform the main Faraday update:
                \begin{equation}
                  \AtN{B^i}{n-1/2} \xRightarrow[\Delta t]{\AtN{E^i}{n}} \AtN{B^i}{n+1/2}
                \end{equation}

        \end{itemize}

  \item Perform the update of $D^i$ field via two Amp\`ere substeps:
        \begin{itemize}
          \item Perform an auxiliary Amp\`ere substep:
                \begin{equation}
                  \AtN{D^i}{n-1/2} \xRightarrow[\Delta t]{~\AtN{H_i}{n},~\AtN{\mathcal{J}^i}{n}} \AtN{D^i}{n+1/2}.
                \end{equation}

          \item Recover auxiliary field $H_i$ at $\AtN{t}{n+1/2}$:
                \begin{equation}
                  \{\AtN{B^i}{n+1/2},~\AtN{D^i}{n+1/2}\} \rightarrow \AtN{H_i}{n+1/2}
                \end{equation}

          \item Perform the final Amp\`ere substep:
                \begin{equation}
                  \AtN{D^i}{n} \xRightarrow[\Delta t]{\AtN{H_i}{n+1/2},~\AtN{\mathcal{J}^i}{n+1/2}} \AtN{D^i}{n+1}.
                \end{equation}
        \end{itemize}

  \item Finally at $\AtN{t}{n+1}$ we have the following quantities:
        \begin{equation*}
          \begin{aligned}
             & \AtN{B^i}{n+1/2},~\AtN{D^i}{n+1};                       \\
             & \AtN{B^i}{n-1/2},~\AtN{D^i}{n};                         \\
             & \AtN{H_i}{n+1/2},~\AtN{E_i}{n+1};                       \\
             & \AtN{\mathcal{J}^i}{n-1/2},~\AtN{\mathcal{J}^i}{n+1/2}; \\
             & \AtN{u_i}{n+1/2},~\AtN{x^i}{n+1},~\AtN{x^i}{n},
          \end{aligned}
        \end{equation*}
        \noindent and the loop is repeated.

\end{enumerate}

\subsection{Comment on the performance}
Many of the GR routines are shared with the main special-relativistic module of the \entity~described in detail in paper I. For example, the current-deposition routine for the conformal electric currents is identical. The field-evolution algorithm is also the same, except that in GR it is executed twice per timestep and additionally includes the recovery step for the auxiliary fields. The principal differences from the performance standpoint thus lies in the particle pusher, and in the amount of data -- for both the particles and fields -- that must be communicated between the meshblocks in multi-GPU runs. In terms of the memory footprint, GR module also stores three times as many electromagnetic-field components (incl. the auxiliary fields, the additional fields at staggered timesteps as well as the electric currents at a previous time).

We benchmarked each major algorithmic component on a single NVIDIA A100 GPU using the monopole test (Sec.~\ref{sec:monopole}) performed with a $2048^2$ mesh which is logarithmic in $r$ and linear in $\theta$ hosting 90 particles per cell. With a well-balanced load, the particle pusher dominates the timestep at about $6.2$ ns per particle; the field solver requires $2\dots3$ ns per cell each step (which is typically negligible with high number of particles per cell). For reference, in the 2D SR tests of paper I, the pusher and current deposition each cost $\approx 1\dots2$ ns per particle, with the pusher again being the most expensive kernel. Communication costs on multi-GPU runs are dominated by particles and are thus identical to those reported in paper I (since the particle arrays are identical). The current deposition cost is identical to that in SR.

\section{Tests}\label{sec:tests}
In this section, we first describe the tests used to validate each routine of the scheme individually. We then present test simulations of the full PIC scheme, comparing our results with both previous studies in the literature and analytic solutions available in the force-free regime.

\subsection{Pusher}
\begin{table*}
  \hspace{-3.5cm}
  \begin{tabular}{c||cccccccccccc }


    Name     & field      & $a$   & $r_{\rm max}/r_g$ & $N$  & $\Delta t / (r_g)$ & $\mathcal{E}$ & $\mathcal{L}/mc$ & $r/r_g$ & $\theta$    & $u_1$ & $u_3$ \\
    \hline
    \hline
    GR       & zero-field & 0.995 & 15                & 512  & 0.01               & 0.92025       & 2                & 10.6498 & $\pi/2$     & 0.18915 & 2.0   \\
    SR-EM    & Wald $a=0$ & --    & 20                & 1024 & 0.01               & 1.41421       & 10.0             & 10.0    & $\pi/2$     & 0     & 10.0  \\
    GR-EM-2D & Wald $a=0$ & 0.0   & 20                & 1024 & 0.01               & 0.873         & 4.5              & 4       & $\pi/2$     & 0.873 & 2.9   \\
    GR-EM-3D & Wald       & 0.9   & 8                 & 1024 & 0.001              & 2.2226        & 18.10399           & 4       & $\pi/2-0.2$ & 0.5664 & 2.361 \\
  \end{tabular}
  \caption{Summary of the test-particle orbits shown in Figure~\ref{fig:orbits}.
    The simulation setup is specified by the field configuration, black hole spin $a$, outer boundary position, and the number of grid cells (identical in the $r$ and $\theta$ directions in all cases). The normalized components of the four-velocity are listed in Kerr-Schild coordinates, with $u_2 = 0$ for all cases.}
  \label{tab:orbits}
\end{table*}

\begin{figure*}
  \centering
  \includegraphics[width=\textwidth]{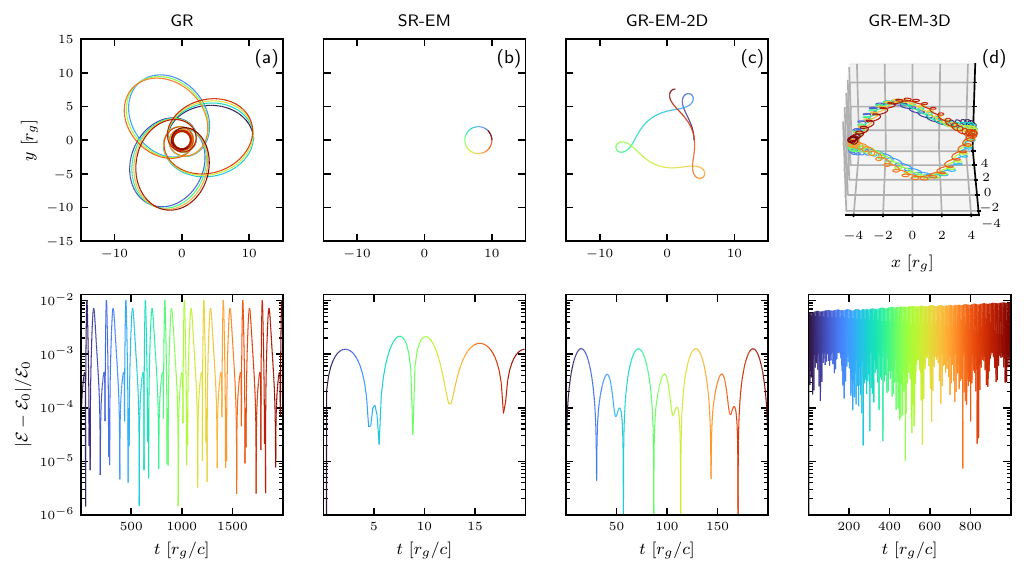}
  \caption{Test particles' orbits (top row) and their corresponding relative energy errors (bottom). }
  \label{fig:orbits}
\end{figure*}
To validate the particle pusher, we separately test its geodesic and electromagnetic components, and then test the full scheme by reproducing results reported in the literature. Table~\ref{tab:orbits} summarizes the orbits considered in this subsection, specifying the magnetic field configuration, the BH spin, and the initial orbital parameters (including the energy $\mathcal{E}$, the angular momentum $\mathcal{L}$, the coordinates $r$ and $\theta$, and the initial velocity components). All the orbits are shown in Figure~\ref{fig:orbits} (top row) with their respective relative errors shown in the bottom row.

Initial conditions of all particles for the tests presented below are characterized by $\mathcal{L}$, the condition that $u_{1, {\rm BL}} = u_{2, {\rm BL}}=0$ (where the subscript ``BL'' means the velocities measured in Boyer-Lindquist coordinates, whereas ``KS'' will denote those in Kerr-Schild coordinates\footnote{We make this distinctions, since most test orbits in the literature are presented in Boyer–Lindquist coordinates, while \entity~operates in Kerr-Schild coordinates.}), and the $\theta$ coordinate. From these quantities, the values for $u_{0, {\rm BL}} = u_{0, {\rm KS}}$ and $\mathcal{E}$ can be deduced. For comparison with our Kerr–Schild framework, the corresponding covariant velocity components can be found using the following:
\begin{align}
  \begin{aligned}
    u_{0,{\rm KS}} & = u_{0,{\rm BL}},                                                               \\
    u_{1,{\rm KS}} & = -\frac{2ru_{0,{\rm BL}} + au_{3,{\rm BL}} }{r^2 - 2r + a^2} + u_{1,{\rm BL}}, \\
    u_{2,{\rm KS}} & = u_{2,{\rm BL}},                                                               \\
    u_{3,{\rm KS}} & = u_{3,{\rm BL}}.
  \end{aligned}\label{eq:BL-KS-transformation}
\end{align}

\noindent The coordinate positions themselves remain unchanged. Below, we ignore the ``KS'' subscript for brevity, implying that all the quantities are in Kerr-Schild coordinates.


The initial radial coordinate is determined from the energy $\mathcal{E}$ via the normalization of the four-momentum for massive particles,
\begin{equation}
  u_\mu u^\mu = g^{\mu \nu} u_\mu u_\nu = -1,
\end{equation}

\noindent which yields a relation via quadratic equation between the particle's coordinates, velocities, and its energy in KS coordinates:
\begin{equation}
  \mathcal{E}=-u_0 = \frac{-B - \sqrt{B^2 - 4 AC}}{2A},
\end{equation}

\noindent where $A \equiv  g^{00}$, $B\equiv -2g^{01}u_1$, and $C \equiv  g^{11} (u_1)^2 + g^{22} (u_2)^2 + g^{33} (u_3)^2 + 2 g^{13} u_1u_3 + 1$. Here the sign is chosen such that $\mathcal{E}>0$. To assess numerical accuracy over time, we monitor the relative error in energy, $|\mathcal{E}(t)/\mathcal{E}_0 -1|$, where $\mathcal{E}_0$ is the initial orbital energy. Small discrepancies with published test-particle trajectories can arise even for apparently identical initial conditions because our integrator is time-staggered: coordinates and velocities are offset by half a timestep. For this reason, when evaluating $\mathcal{E}(t)$ we align variables by taking the position at the velocity time level, i.e., the midpoint (average) of the position at the adjacent time steps.

\subsubsection{Geodesics force}
We first reproduce the three-leaf orbit described in \cite{Levin_LeafOrbits2008PhRvD} (see their Fig.~15, also reproduced in \citealt{Zou2025PhRvD.111h3032Z}) as a demonstration of our geodesic pusher (Fig.~\ref{fig:orbits}a). We use a timestep of $\Delta t = 0.01 ~r_g$ and evolve the orbit up to $2000 ~r_g$. We successfully reproduce the orbit, which exhibits a small precession, and while the energy error grows as the particle approaches pericenter. The error remains stable and bounded over long timescales (Fig.~\ref{fig:orbits}a bottom).

\subsubsection{Lorentz force}
We now turn to orbits in the presence of a magnetic field. The Wald solution \citep{Wald1974PhRvD..10.1680W} describes an electromagnetic four-potential corresponding to a magnetic field aligned with the BH spin axis that is uniform at spatial infinity. In Kerr–Schild coordinates, the vector potential $A^\mu$ defining the zero-charge solution is:
\begin{equation}
  \begin{split}
    A_0 = \frac{B_0}{2} \left( h_{13} \beta^1 + 2 a g_{00} \right), \\
    A_1 = \frac{B_0}{2} \left( h_{13} + 2 a h_{11} \beta^1\right),  \\
    A_3 = \frac{B_0}{2} \left( h_{33} + 2a h_{13} \beta^1\right),
  \end{split}\label{eq:Wald}
\end{equation}
where $B_0$ is the asymptotic magnetic field strength. To set this field in the code, we compute $B^i$ according to Eq.~\eqref{eq:EB-set}, and then compute gravitationally-induced $D^i$ in KS coordinates.

To test the electromagnetic pusher in isolation, we consider the general-relativistic metric with flat space-time (no gravity, and $a=0$), which is mathematically equivalent to a spherical grid modeled using the SR module (for convenience, we still measure temporal and spatial dimensions in $r_g$ which in this case bears no physical meaning, i.e., the metric does not depend on its value). In this case, the magnetic field reduces to a uniform vertical field (in the $r$–$\theta$ plane). This setup allows us to test simple particle gyration in a nontrivial coordinate system, requiring transformations between covariant components and the tetrad frame. The orbit, shown in the second column of Fig.~\ref{fig:orbits}, is initialized in a vertical field with fiducial Larmor radius $\rho_0=2 ~r_g$ (see paper I for details on code units and magnetic field normalization), using a timestep $\Delta t = 0.01~ r_g$.

This setup corresponds to a physical Larmor radius of the particle of $\rho = \rho_0 u_{\hat \phi} = \rho_0 u_{\phi} / r = 2~r_g$, as illustrated in Fig.\ref{fig:orbits}b. The trajectory remains closed and the energy error stays bounded (Fig.~\ref{fig:orbits}b bottom).

\subsubsection{Full pusher}
We now test the full pusher scheme for an orbit in the $r$–$\phi$ plane for the nonspinning case ($a=0$). The results of numerical integration are shown in Fig.\ref{fig:orbits}c. We initialize the magnetic field according to the Wald’s solution, as given by Eq.~\eqref{eq:Wald}, which corresponds to a purely vertical uniform magnetic field in Boyer–Lindquist coordinates. However, in Kerr-Schild coordinates, a nonzero electric field arises from the coupling of the shift $\beta^1$ with $B^2$. As a benchmark, we reproduce one of the charged-particle orbits reported in \cite{ChargedPrtcl2015CQGra}. For a vertical field, the conjugate angular momentum includes the magnetic field contribution; hence, for a non-spinning BH, the toroidal velocity can be inferred from
\begin{equation}
  u_\phi = \mathcal{L} - mr^2 \sin^2 \theta \frac{qB}{2m}.
\end{equation}

We reproduce the orbit presented in Figure 6.2 of \cite{ChargedPrtcl2015CQGra}. To match the magnetic field normalization for $\mathcal{B} = qB/2m$, we adopt the fiducial Larmor radius $\rho_0 = (q/m)/(2 |\mathcal{B}|)$, corresponding to $\rho_0 = 5~r_g$ for $q/m=1$ and $|\mathcal{B}| = 0.1$.

Following \citet{Stuchlik2016EPJC...76...32S}, we validate fully three-dimensional test-particle trajectories in the analytic vacuum Wald magnetic field (Eq.~\eqref{eq:Wald}) for a Kerr black hole with a dimensionless spin $a=0.9$ (with the magnetic field aligned with the spin axis). Figure~\ref{fig:orbits}d reproduces Fig.~11 (right column) from the same authors, showing a regular (non-chaotic) orbit. We adopt their normalization $\mathcal{B}=1$, corresponding to a fiducial Larmor radius $\rho_0=0.5~r_g$.
  {To initialize the orbit, we treat the particle as neutral on a Keplerian trajectory: we take $u_t$ and $u_\phi$ from Eqs.~(63–64) of \citet{Stuchlik2016EPJC...76...32S} and obtain the corresponding Kerr–Schild radial component $u_r$ via Eq.~\eqref{eq:BL-KS-transformation}.} Note that similar orbits were studied by \citet{Bacchini2019ApJS} and \citet{APERTURE}, albeit with different initial velocities. In contrast to the other orbits discussed above, we observe a slow, monotonic drift in the relative energy error. However, the error remains small and bounded. This secular trend is due to the interpolation of the spatially varying magnetic field (see also \citealt{Bacchini2019ApJS}).

\subsection{Wald solution in vacuum}
\begin{figure*}
  \centering
  \includegraphics[width=\textwidth]{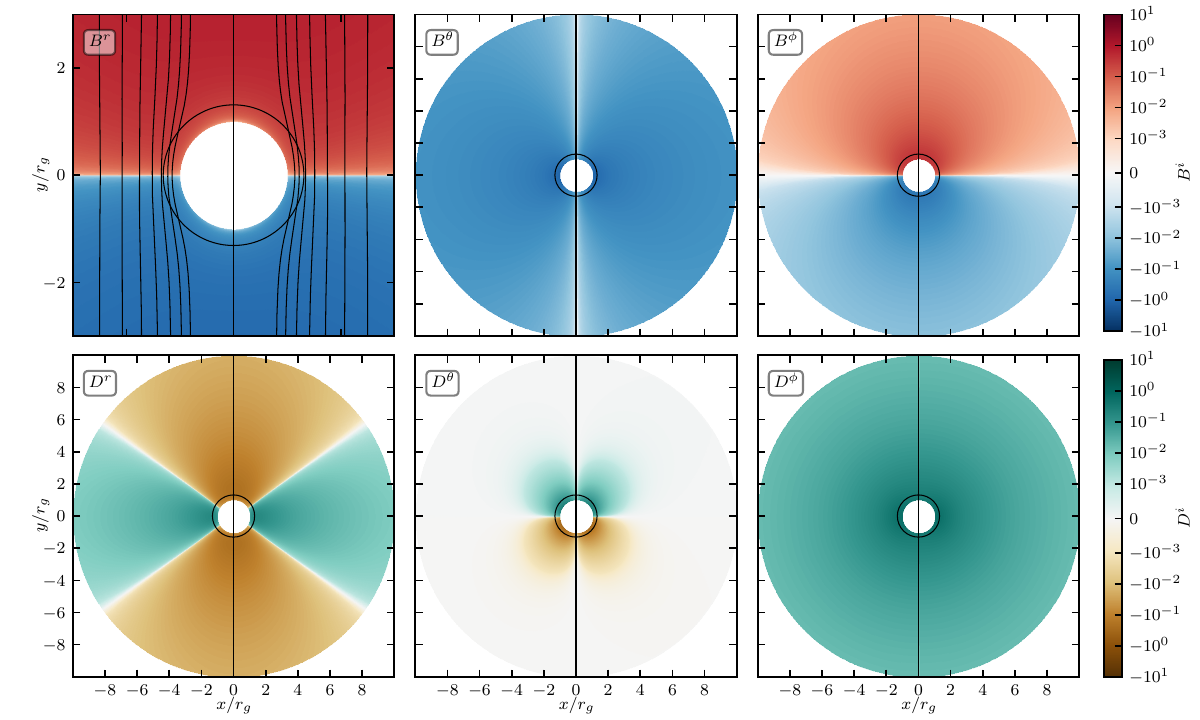}
  \caption{Conservation of Wald solution in vacuum, from the initial state (left side of each panel) to $t=250~r_g$ (right side of each panel). The top row shows the magnetic field components $B^i$, while the bottom row shows the electric field components $D^i$. In the first panel, black contours trace the magnetic field lines and we show a zoom-in close to the BH. In all panels, the event horizon is indicated by a thin, round black outline.}
  \label{fig:vacuum}
\end{figure*}
We now test the evolution of electromagnetic fields in vacuum, i.e., in the absence of plasma. The problem generator for this test is included in \entity, with two available initial field configurations defined in the input file via the \texttt{init\_field} setting, which can be set to either \texttt{wald} or \texttt{vertical}. The former initializes the fields according to Eq.~\eqref{eq:Wald}, while the latter corresponds to a purely vertical magnetic field -- the zero-spin limit of Eq.~\eqref{eq:Wald}.

Figure~\ref{fig:vacuum} shows the evolution of the Wald solution in vacuum for a spin parameter $a=0.95$. Each of the six panels represents one component of the electromagnetic field: the magnetic field ($B^r$, $B^\theta$, $B^\phi$) in the top row and the electric field ($D^r$, $D^\theta$, $D^\phi$) in the bottom row. In each panel, the left half ($x<0$) displays the initial condition, while the right half ($x>0$) shows the state at $t=500r_g$. In the first panel, thin black lines overlay $B^r$ to illustrate the magnetic field lines. The simulation was performed at a resolution of $512^2$ with a grid uniform in $\theta$ and logarithmic in $r$. In the absorbing layer, we smoothly match the field components with the analytic solution. The solution remains stable and well preserved after $\sim 10^5$ timesteps, matching exactly the analytic expression.

\subsection{Deposit}
\begin{figure*}
  \centering
  \includegraphics[width=\textwidth]{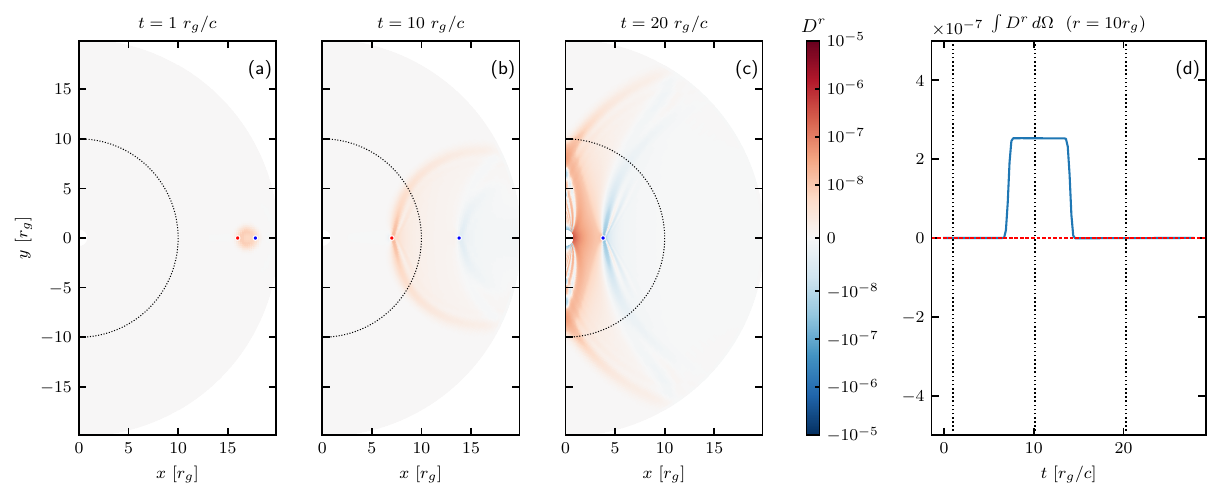}
  \caption{Panels (a-c): Snapshots of the radial component of the electric field $D^r$ (in color) at three different times
    ($t = 1\,r_g$, $10\,r_g$, and $20\,r_g$). The dotted semicircle indicates the reference radius $r = 10\,r_g$.
    The blue and red markers denote the two particles (electron and positron).
    Panel (d): Time evolution of the angular integral $\int D^r\,d\Omega$ at $r = 10\,r_g$, with vertical dashed lines marking the snapshot times.}
  \label{fig:deposit}
\end{figure*}
We next test the current deposition scheme in the presence of a rapidly spinning black hole ($a=0.95$). The system is initialized without any electromagnetic fields, and an electron–positron pair is placed at $(r,\theta) = (17~r_g, \pi/2)$. The particles are given equal and opposite radial velocities, $u_r = +5$ for the electron and $u_r = -5$ for the positron. At the outer boundary, we impose reflection for the electron instead of absorption. As a result, both particles eventually fall into the BH, as illustrated in the snapshots of Figure~\ref{fig:deposit}a–c, which show the particle trajectories superimposed on the evolving electric field distribution.

Panel (d) of Figure~\ref{fig:deposit} shows the evolution of the flux of $D^r$, computed over a spherical surface at $r=10~r_g$. The integral initially increases as the particles move, but ultimately decays to zero, demonstrating perfect charge conservation. 

\subsection{Plasma-filled magnetosphere}
\begin{figure}
  \centering
  \includegraphics[width=\linewidth]{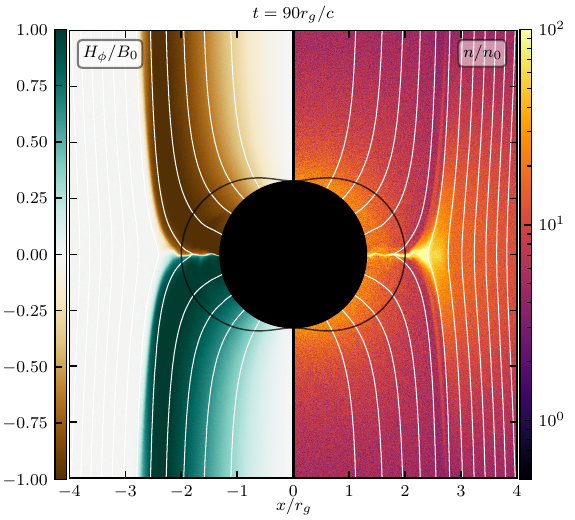}
  \caption{Snapshot of the plasma-filled magnetosphere around a rapidly rotating black hole ($a = 0.95$) at $t = 90\,r_g$. The color scales show $H_\phi$ normalized to the fiducial $B_0$ (left), and the plasma number density normalized to the fiducial density $n_0$ (right). White curves indicate magnetic field lines. The black circle marks the event horizon, and the vertical solid line denotes the rotation axis. Continuous pair injection maintains densities at or above the Goldreich–Julian level, resulting in a filled magnetosphere consistent with the high-supply regime of \cite{Parfrey2019PhRvL}.}
  \label{fig:accretion}
\end{figure}
To validate the full PIC scheme, we reproduce the setup of \cite{Parfrey2019PhRvL}. Specifically, we initialize the Wald solution corresponding to a uniform vertical magnetic field around a rapidly spinning black hole with spin parameter $a = 0.95$, and continuously inject plasma into the magnetosphere to replenish the lost plasma. Our computational domain extends to $6r_g$, with an absorbing layer beginning at $5r_g$. We use a grid of $2048^2$ cells with logarithmic spacing in $r$. As plasma accretes, both the density and magnetic field strength increase, and the steady-state skin depth near the black hole is established at $\approx0.005~r_g$ (the fiducial skin depth for $n=n_0$ is taken to be $d_0=0.05~r_g$, while the fiducial Larmor radius $\rho_0=0.025~r_g$). The nominal number of particles per cell is $\mathtt{nppc}=2$, but the effective value is substantially higher owing to continuous plasma injection. Electron–positron pairs are injected with a temperature $T_\pm = 0.5 m_\pm$ whenever local magnetization exceeds $\sigma\equiv B_iB^i/(4\pi n_\pm m_\pm) \geq 1000$. This setup corresponds to the standard \texttt{accretion} problem generator provided with \entity.

As the simulation proceeds, the magnetosphere rapidly becomes plasma-loaded. Figure~\ref{fig:accretion} shows the state at $t=90~ r_g$. With increasing plasma supply, the field lines deviate from the vacuum solution and bend toward the BH near and within the ergosphere. After roughly $10~ r_g$, the system settles into a quasi–steady state characterized by an approximately monopolar poloidal field inside the ergosphere, and a persistent Y-point and current sheet. In this regime, the current sheet reconnects continuously and is susceptible to a drift–kink instability. The resulting plasma distribution exhibits the expected high-supply structure, consistent with \cite{Parfrey2019PhRvL}.

\subsection{BZ monopole}\label{sec:monopole}
\begin{figure*}
  \centering
  \includegraphics[width=\textwidth]{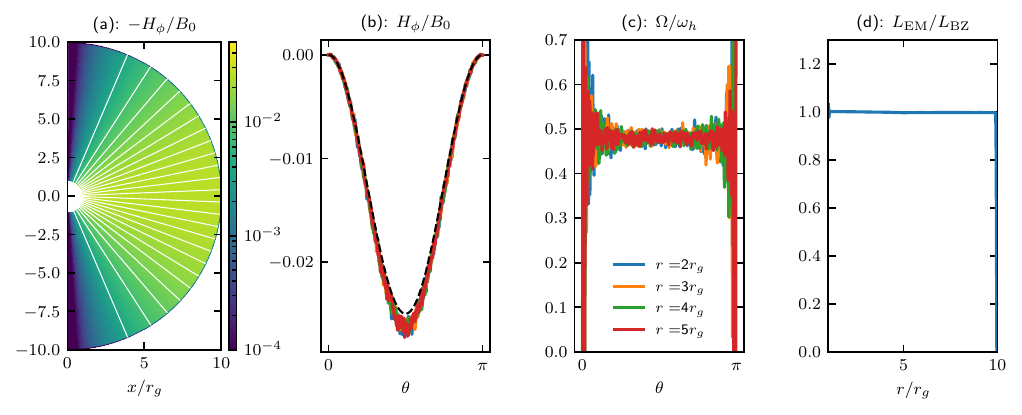}
  \caption{Snapshot of the Blandford–Znajek monopole solution in GRPIC at $t = 200r_g$ for a BH spin $a = 0.2$. Panel (a) shows the distribution of $H_\phi/B_0$ across the domain, with white lines indicating magnetic field lines. Panels (b) and (c) present the $\theta$-profiles of $H_\phi/B_0$ and $\Omega/\omega_h$, respectively, at four different radii; the analytic solution for $H_\phi$ is shown by the black dashed line in panel (b). Panel (d) displays the electromagnetic Poynting flux as a function of radius, normalized by the Blandford–Znajek luminosity, $L_{\rm BZ}$.
  }
  \label{fig:BZ}
\end{figure*}
In this test we initialize a zero-spin magnetic monopole solution centered on the black hole, reproducing the classic BZ solution (\cite{BlandfordZnajek1977MNRAS.179..433B}; see also \citealt{Komissarov2001MNRAS.326L..41K, KomissarovMonopole2004MNRAS.350.1431K} for force-free and GRMHD extensions):
\begin{equation}
  A_\phi = - B_0 \cos \theta,
\end{equation}
which yields $B^r = B_0 \sin \theta / \sqrt{h}$, with all other field components set to zero. The computational domain extends to $r_{\rm max}=10~r_g$, with an absorbing layer of $1~r_g$ thick near the outer boundary. The resolution is $1024^2$, and the black hole spin is set to $a=0.2$ (though we have also tested higher spins). The inner boundary lies at $r_{\rm min}=r_g$.

We adopt fiducial skin depths $d_0=2.5\cdot 10^{-3}r_g$ and $\rho_0 = 5\cdot 10^{-5}r_g$, resulting in the value for the fiducial plasma magnetization of $\sigma_0 = 2500$. Maxwellian plasma with $T_\pm =0.5 m_\pm$ is injected wherever the local magnetization exceeds $\sigma_{\rm th} = 2000$. The density profile follows $n \propto r^{-2}$ scaling. The choice of $\sigma_{\rm th}$ yields densities significantly above the GJ value, leading to high plasma multiplicity. The resulting $H_\phi$ at $200r_g$ is shown in Fig.~\ref{fig:BZ}a.

To test convergence toward the BZ solution, we compute $H_\phi$ and compare it with the analytic monopole solution of \citet{BlandfordZnajek1977MNRAS.179..433B}:
\begin{equation}
  H_\phi = -B_0 \frac{a \sin^2 \theta}{8} + O(a^3),
\end{equation}
at several radii (Fig.~\ref{fig:BZ}b), showing good agreement between simulation results and BZ calculation.

We further compute the angular velocity of magnetic field lines as
\begin{equation}
  \Omega = - \frac{E_\theta}{\sqrt{h} B^r},
\end{equation}
and find that it peaks at the expected value of $0.5\omega_h$, where $\omega_h$ is the black hole angular frequency (given below).

Finally, we evaluate the electromagnetic Poynting flux,
\begin{equation}
  L_{\rm EM} = 2 \pi \int_0^{\pi} P^r \sqrt{h} d\theta,
\end{equation}
where
\begin{equation}
  P^r = \frac{1}{4 \pi \sqrt{h}} \left( E_\theta H_\phi - E_\phi H_\theta \right),
\end{equation}
and find that it roughly follows the analytic BZ luminosity (Fig.~\ref{fig:BZ}d),
\begin{align}
  \begin{aligned}
    L_{\rm BZ} = \frac{B_0^2 \omega_h^2}{6}, \\
    \text{where}                             \\
    \omega_h = - \frac{g_{03}(r_h)}{g_{33}(r_h)} = \frac{a}{r_h}.
  \end{aligned}
\end{align}

\section{Summary}

In this paper, we presented the general relativistic module of the architecture-agnostic open-source PIC code -- \entity. The module implements a set of algorithms which integrate the trajectories of charged particles and the electromagnetic fields in the axisymmetric (2.5D) curved spacetime using the $3{+}1$ formalism in Kerr–Schild coordinates, with an optional quasi-spherical mapping. We demonstrated a set of standard problem generators and tests, together with the relevant parameter inputs, both of which are distributed with the core of the code and are freely available\footnote{GitHub link: \href{https://github.com/entity-toolkit/entity}{github.com/entity-toolkit/entity}.}. While the present release supports 2.5D axisymmetric Kerr-Schild coordinates, future work will extend the method to 3D using the cubed-sphere mesh.


\begin{acknowledgments}
  Authors would also like to thank K. Parfrey and E. Quataert for their helpful comments and countless fruitful discussions. The initial development of the code was supported by U.S. DOE grant DE-AC02-09CH11466, NSF Cyberinfrastructure for Sustained Scientific Innovation grant 2311800, and the NVIDIA Corporation Academic Hardware Grant Program. The authors would also like to acknowledge the OLCF Director's Discretion Project AST214 which enabled the testing of the code on the \texttt{Frontier} supercomputer. This work was facilitated by Simons Foundation (00001470, A.P.), and Multimessenger Plasma Physics Center (MPPC, A.P.), NSF grant No. PHY-2206607. A.P. additionally acknowledges support by an Alfred P. Sloan Fellowship, and a Packard Foundation Fellowship in Science and Engineering. The computations in this work were, in part, run at facilities supported by the Scientific Computing Core at the Flatiron Institute, a division of the Simons Foundation. The developers are pleased to acknowledge that the work was performed using the Princeton Research Computing resources at Princeton University which is a consortium of groups led by the Princeton Institute for Computational Science and Engineering (PICSciE) and Office of Information Technology's Research Computing.
\end{acknowledgments}

\appendix

\section{Metric}\label{ap:metric}
We adopt Kerr–Schild (KS) coordinates throughout (see the project wiki for reference formulas). The expressions below provide all metric components, tetrads, and spatial derivatives required by the GRPIC particle pusher. In practice, the $3{+}1$ split supplies the lapse, shift, and spatial metric for field evolution, while metric derivatives enter the particle pusher for accurate trajectory integration. Tetrad transformations provide a locally orthonormal frame that the pusher uses for electromagnetic fields and particle velocities.

The covariant and contravariant metrics are
\begin{equation}
  g_{\mu \nu}=\begin{bmatrix}-(1-z) & z & \cdot & -az\sin^2{\theta}\\z & 1+z & \cdot & -a(1+z)\sin^2{\theta} \\\cdot & \cdot & \Sigma & \cdot \\-az\sin^2{\theta} & -a(1+z)\sin^2{\theta} & \cdot & \frac{A\sin^2{\theta}}{\Sigma}\end{bmatrix},~~~
  g^{\mu\nu}=\begin{bmatrix}-(1+z) & z & \cdot & \cdot\\z & \frac{\Delta}{\Sigma} & \cdot & \frac{a}{\Sigma}\\\cdot & \cdot & \frac{1}{\Sigma} & \cdot\\\cdot & \frac{a}{\Sigma} & \cdot & \frac{1}{\Sigma\sin^2{\theta}}\end{bmatrix}
\end{equation}
with
\begin{align*}
   & \Sigma=r^2+a^2\cos^2\theta, \qquad
  \Delta = r^2-2r+a^2, \qquad
  A=(r^2+a^2)^2-a^2\Delta \sin^2\theta, \\
   & z=\dfrac{2r}{\Sigma}, \qquad
  \sqrt{-g}=\Sigma\sin\theta.
\end{align*}

In the $3{+}1$ decomposition, the spatial metric and its inverse are
\begin{equation}
  h_{ij}=
  \begin{bmatrix}
    1+z                 & \cdot  & -a(1+z)\sin^2\theta           \\
    \cdot               & \Sigma & \cdot                         \\
    -a(1+z)\sin^2\theta & \cdot  & \dfrac{A\sin^2\theta}{\Sigma}
  \end{bmatrix},
  \quad
  h^{ij}=
  \begin{bmatrix}
    \dfrac{A}{\Sigma(\Sigma + 2r)} & \cdot             & \dfrac{a}{\Sigma}             \\
    \cdot                          & \dfrac{1}{\Sigma} & \cdot                         \\
    \dfrac{a}{\Sigma}              & \cdot             & \dfrac{1}{\Sigma\sin^2\theta}
  \end{bmatrix}.
\end{equation}
The lapse function $\alpha$ and shift vector $\beta^i$ are
\begin{align*}
  \beta^i=\begin{bmatrix}\dfrac{z}{1+z}\\ \cdot \\ \cdot\end{bmatrix},
  \qquad
  \alpha^2=\dfrac{1}{1+z},
  \qquad
  \beta_i=\bigl[z~~\cdot~~-a z \sin^2\theta\bigr],
  \qquad
  \beta_i\beta^i=\dfrac{z^2}{1+z},
\end{align*}
with $\sqrt{h}=\Sigma\sin\theta/\alpha$. Thus, specifying $\beta^1$ is sufficient.

It is convenient to define a locally orthonormal tetrad basis, constructed via
\[
  e^i_{\ \hat{i}}=
  \begin{bmatrix}
    \sqrt{h^{rr}}                          & \cdot                     & \cdot                 \\
    \cdot                                  & 1/\sqrt{h_{\theta\theta}} & \cdot                 \\
    -\sqrt{h^{rr}}\,h_{r\phi}/h_{\phi\phi} & \cdot                     & 1/\sqrt{h_{\phi\phi}}
  \end{bmatrix},
  \qquad
  e_i^{\ \hat{i}}=
  \begin{bmatrix}
    1/\sqrt{h^{rr}} & \cdot                   & h_{r\phi}/\sqrt{h_{\phi\phi}} \\
    \cdot           & \sqrt{h_{\theta\theta}} & \cdot                         \\
    \cdot           & \cdot                   & \sqrt{h_{\phi\phi}}
  \end{bmatrix}.
\]
Transformations between coordinate and tetrad bases follow
\[
  A^{\hat{i}}=e^{\hat{i}}_{\ j} A^j,\qquad
  A^{i}=e_{\ \hat{j}}^{i} A^{\hat{j}},\qquad
  a_{\hat{i}}=e_{\hat{i}}^{\ j} a_{j},\qquad
  a_{i}=e_{i}^{\ \hat{j}} a_{\hat{j}}.
\]
Explicitly,
\begin{align*}
   & A^{\hat{r}}=\frac{1}{\sqrt{h^{rr}}}A^r,
   &                                                                                    & a_{\hat{r}}=\sqrt{h^{rr}}\,a_r-\sqrt{h^{rr}}\frac{h_{r\phi}}{h_{\phi\phi}}\,a_\phi, \\
   & A^{\hat{\theta}}=\sqrt{h_{\theta\theta}}\,A^\theta,
   &                                                                                    & a_{\hat{\theta}}=\frac{1}{\sqrt{h_{\theta\theta}}}\,a_\theta,                       \\
   & A^{\hat{\phi}}=\frac{h_{r\phi}}{\sqrt{h_{\phi\phi}}}A^r+\sqrt{h_{\phi\phi}}A^\phi,
   &                                                                                    & a_{\hat{\phi}}=\frac{1}{\sqrt{h_{\phi\phi}}}\,a_\phi,
\end{align*}
and the inverse,
\begin{align*}
   & A^{r}=\sqrt{h^{rr}}\,A^{\hat{r}},
   &                                                                                                         & a_{r}= \frac{1}{\sqrt{h^{rr}}}\,a_{\hat{r}} + \frac{h_{r\phi}}{h_{\phi\phi}}\,a_{\hat{\phi}}, \\
   & A^{\theta}=\frac{1}{\sqrt{h_{\theta\theta}}}\,A^{\hat{\theta}},
   &                                                                                                         & a_{\theta}=\sqrt{h_{\theta\theta}}\,a_{\hat{\theta}},                                         \\
   & A^{\phi}=-\frac{h_{r\phi} h^{rr}}{h_{\phi\phi}}A^{\hat{r}}+\frac{1}{\sqrt{h_{\phi\phi}}}A^{\hat{\phi}},
   &                                                                                                         & a_{\phi}=\sqrt{h_{\phi\phi}}\,a_{\hat{\phi}}.
\end{align*}

Assuming general parameterization of coordinates, $\theta=\theta(y)$ and $r=r(x)$, the spatial derivatives required for the pusher algorithm are
\begin{align*}
   & \partial_y h^{11} = \frac{\Sigma (\Sigma + 2r)\,\partial_y A - 2A\,\partial_y\Sigma\,(\Sigma + r)}{\Sigma^2(\Sigma+2r)^2},
   &                                                                                                                                                & \partial_y h^{22} = -\frac{\partial_y \Sigma}{\Sigma^2},               \\
   & \partial_y h^{33} = - \frac{\partial_y \Sigma + 2 \frac{\cos\theta}{\sin\theta}\,\Sigma\,\partial_y \theta}{\left(\Sigma \sin\theta\right)^2},
   &                                                                                                                                                & \partial_y h^{13} = - \frac{a\, \partial_y \Sigma}{ \Sigma^2},         \\
   & \partial_y A = -2 a^2 \Delta \sin\theta \cos\theta\, \partial_y \theta,
   &                                                                                                                                                & \partial_y \Sigma = -2a^2 \cos\theta \sin\theta\, \partial_y \theta,   \\
   & \partial_y \beta^1 = - \frac{2 r\, \partial_y \Sigma}{\Sigma^2 (1 + z)^2}\, ,                                                                  &                                                                      &
\end{align*}
and
\begin{align*}
   & \partial_x h^{11} = \frac{\Sigma (\Sigma + 2 r)\,\partial_x A -
    2A \bigl(r\,\partial_x \Sigma + \Sigma(\partial_x \Sigma + \partial_x r)\bigr)}{\Sigma^2(\Sigma + 2 r)^2},
   &                                                                                          & \partial_x h^{22} = -\frac{\partial_x \Sigma}{\Sigma^2},                                                  \\
   & \partial_x h^{33} =  - \frac{\partial_x \Sigma}{\Sigma^2 \sin^2 \theta},
   &                                                                                          & \partial_x h^{13} = - \frac{a\, \partial_x \Sigma }{\Sigma^2},                                            \\
   & \partial_x A =  4 r\,\partial_x r\, (r^2 + a^2) - a^2 \sin^2 \theta\, \partial_x \Delta,
   &                                                                                          & \partial_x \Sigma = 2r\,\partial_x r,                                                                     \\
   & \partial_x \Delta = 2(r - 1)\,\partial_x r,
   &                                                                                          & \partial_x \beta^1 = \frac{2\bigl(\partial_x r\,\Sigma - r\,\partial_x \Sigma\bigr)}{(\Sigma + 2r)^2}\, .
\end{align*}

For each of the three metrics compatible with the GRPIC module in \entity, we explicitly provide the covariant and contravariant 3-metric components, tetrad transformations, and metric derivatives required for the particle pusher.

\subsection{Spherical Minkowski}
For all of our 2D simulations we adopt $d\phi \equiv 1$. The \texttt{kerr\_schild\_0} metric in \entity\ is used primarily for testing purposes; it corresponds to a spherical grid in flat Minkowski spacetime. The spatial metric and its inverse are
\begin{equation}
  h_{ij}=
  \begin{bmatrix}
    (dr)^2 & \cdot           & \cdot                      \\
    \cdot  & (d\theta)^2 r^2 & \cdot                      \\
    \cdot  & \cdot           & (d\phi)^2 r^2 \sin^2\theta
  \end{bmatrix},
  \quad
  h^{ij}=
  \begin{bmatrix}
    (dr)^{-2} & \cdot                        & \cdot                                  \\
    \cdot     & (d\theta)^{-2}\dfrac{1}{r^2} & \cdot                                  \\
    \cdot     & \cdot                        & (d\phi)^{-2}\dfrac{1}{r^2\sin^2\theta}
  \end{bmatrix}.
\end{equation}
The lapse and shift are trivial:
\[
  \beta^1=0, \qquad \alpha^2=1,
\]
and $\sqrt{h}=dr\, d\theta\, d\phi\, r^2\sin\theta$. The functional form of the polar area $A$ used for the field solver is:
\[
  A = dr \, r^2 \left( 1 - \cos \frac{d \theta}{2} \right) .
\]
All derivatives vanish except
\begin{align*}
  \partial_r h^{33} = -\,dr\,\frac{2}{r^3 \sin^2 \theta}, \qquad
  \partial_r h^{22} = -\,dr\,\frac{2}{r^3 (d\theta)^2}, \qquad
  \partial_\theta h^{33} =  -\,2\,d\theta\, \frac{\cos \theta}{r^2 \sin^3 \theta}.
\end{align*}

\subsection{Spherical Kerr-Schild}
In spherical Kerr–Schild coordinates, \texttt{kerr\_schild}, the spatial metric and its inverse are
\begin{equation}
  h_{ij}=
  \begin{bmatrix}
    (dr)^2 (1+z)                      & \cdot              & -dr\, d\phi\, a(1+z)\sin^2\theta        \\
    \cdot                             & (d\theta)^2 \Sigma & \cdot                                   \\
    - dr\, d\phi\, a(1+z)\sin^2\theta & \cdot              & (d\phi)^2 \dfrac{A\sin^2\theta}{\Sigma}
  \end{bmatrix},
  \quad
  h^{ij}=
  \begin{bmatrix}
    (dr)^{-2} \dfrac{A}{\Sigma(\Sigma + 2r)} & \cdot                            & (dr\, d \phi)^{-1} \dfrac{a}{\Sigma}        \\
    \cdot                                    & (d\theta)^{-2} \dfrac{1}{\Sigma} & \cdot                                       \\
    (dr\, d \phi)^{-1}\dfrac{a}{\Sigma}      & \cdot                            & (d \phi)^{-2} \dfrac{1}{\Sigma\sin^2\theta}
  \end{bmatrix}.
\end{equation}
The corresponding lapse function and shift vector are
\[
  \beta^1=(dr)^{-1}\frac{z}{1+z}, \qquad \alpha^2=\frac{1}{1+z},
\]
and $\sqrt{h}=dr\, d\theta\, d\phi\, \Sigma\sin\theta/\alpha$. The functional form of the polar area $A$ used for the field solver is:
\[
  A = dr \left( r^2 + a^2 \right) \sqrt{1 + \frac{2 r}{r^2 + a^2}} \left( 1 - \cos \frac{d \theta}{2} \right) .
\]

The derivatives used by the pusher are
\begin{align*}
   & \partial_r h^{11} = \frac{1}{(dr)^2} \frac{\Sigma (\Sigma + 2 r)\, \partial_r A - 2 A \bigl( r\, \partial_r \Sigma + \Sigma [\partial_r \Sigma + dr] \bigr)}{\Sigma^2 (\Sigma + 2r)^2},
   &                                                                                                                                                                                         & \partial_\theta h^{11} = \frac{1}{(dr)^2} \frac{\Sigma (\Sigma + 2 r)\, \partial_\theta A - 2 A\, \partial_\theta \Sigma\, (r+\Sigma)}{\Sigma^2 (\Sigma + 2r)^2}, \\
   & \partial_r h^{22} = - \frac{1}{(d\theta)^2}\frac{\partial_r \Sigma}{\Sigma^2},
   &                                                                                                                                                                                         & \partial_\theta h^{22} = - \frac{1}{(d\theta)^2}\frac{\partial_\theta \Sigma}{\Sigma^2},                                                                          \\
   & \partial_r h^{33} = -\frac{\partial_r \Sigma}{ \Sigma^2 \sin^2 \theta},
   &                                                                                                                                                                                         & \partial_\theta h^{33} = - \frac{2 \cos\theta\, d\theta}{\Sigma^2 \sin^3\theta} \left( \Sigma - a^2 \sin^2\theta \right),                                         \\
   & \partial_r h^{13} = -\frac{a}{dr}\, \frac{\partial_r \Sigma}{ \Sigma^2},
   &                                                                                                                                                                                         & \partial_\theta h^{13} = -\frac{a}{dr}\, \frac{\partial_\theta \Sigma}{ \Sigma^2},                                                                                \\
   & \partial_r \Delta = 2\,dr\,(r - 1),
   &                                                                                                                                                                                         & \partial_\theta \Delta = 0,                                                                                                                                       \\
   & \partial_r A = 4 r\,dr\, (r^2 + a^2) - a^2 \sin^2 \theta \, \partial_r \Delta,
   &                                                                                                                                                                                         & \partial_\theta A = -2 a^2 \sin \theta \cos \theta \, \Delta \, d\theta,                                                                                          \\
   & \partial_r \Sigma = 2 r\, dr,
   &                                                                                                                                                                                         & \partial_\theta \Sigma = -2 a^2 \sin \theta \cos \theta \, d\theta,                                                                                               \\
   & \partial_r \beta^1 = \frac{2}{dr}\, \frac{dr\, \Sigma - r\, \partial_r \Sigma}{(\Sigma + 2r)^2},
   &                                                                                                                                                                                         & \partial_\theta \beta^1 = -\frac{2}{dr} \frac{r\, \partial_\theta \Sigma}{\Sigma^2(1+z)^2},                                                                       \\
   & \partial_r \alpha = - \frac{dr\, \Sigma - r\, \partial_r \Sigma}{\Sigma^2}\, \alpha^3,
   &                                                                                                                                                                                         & \partial_\theta \alpha = \frac{r\, \partial_\theta \Sigma}{\Sigma^2}\, \alpha^3.
\end{align*}

\subsection{Quasi-spherical Kerr–Schild}
We also implement a quasi-spherical KS coordinate system \texttt{qkerr\_schild}. The radial and polar coordinates are remapped as
\begin{align*}
   & \xi = \log(r - r_0), \qquad
  \theta = \eta + \frac{2 h_0\, \eta (\pi - 2 \eta) (\pi - \eta)}{\pi^2}, \\
   & dr = d\xi\, e^\xi, \qquad
  d\theta = d\eta \left[1 + 2h_0 + 12h\,\frac{\eta}{\pi} \left( \frac{\eta}{\pi} - 1 \right) \right],
\end{align*}
where $r_0$ is a radial offset and $h_0$ controls the concentration of grid points toward the equator.

All spherical KS coefficients remain valid but are now functions of $(\xi,\eta)$. The spatial metric and its inverse are
\begin{subequations}
  \begin{equation}
    h_{ij}=
    \begin{bmatrix}
      (d\xi)^2 e^{2 \xi} (1+z)                    & \cdot                                               & -d\xi\, e^{\xi}\, d\phi\, a(1+z)\sin^2\theta \\
      \cdot                                       & \bigl(d\eta\, \tfrac{d\theta}{d\eta}\bigr)^2 \Sigma & \cdot                                        \\
      - d\xi\, e^\xi\, d\phi\, a(1+z)\sin^2\theta & \cdot                                               & (d\phi)^2 \dfrac{A\sin^2\theta}{\Sigma}
    \end{bmatrix},
  \end{equation}
  \begin{equation}
    h^{ij}=
    \begin{bmatrix}
      (d\xi\, e^{\xi})^{-2} \dfrac{A}{\Sigma(\Sigma + 2r)} & \cdot                                                               & (d\xi\, e^{\xi}\, d \phi)^{-1} \dfrac{a}{\Sigma} \\
      \cdot                                                & \bigl(d\eta\, \tfrac{d \theta}{d \eta}\bigr)^{-2} \dfrac{1}{\Sigma} & \cdot                                            \\
      (d\xi\, e^{\xi}\, d \phi)^{-1}\dfrac{a}{\Sigma}      & \cdot                                                               & (d \phi)^{-2} \dfrac{1}{\Sigma\sin^2\theta}
    \end{bmatrix}.
  \end{equation}
\end{subequations}
The lapse and shift are
\[
  \beta^1= (d\xi\, e^\xi)^{-1}\frac{z}{1+z}, \qquad
  \alpha^2=\frac{1}{1+z},
\]
and $\sqrt{h}=d\xi\, e^\xi\, d\eta\, \bigl(\tfrac{d\theta}{d\eta}\bigr)\, d \phi\, \Sigma\sin\theta/\alpha$. The functional form of the polar area $A$ used for the field solver repeats the functional form of \texttt{kerr\_schild} but expresses it through the $\xi$ and $\eta$ variables in practice.

The derivatives needed by the pusher are
\begin{align*}
                                                                    & \partial_\xi h^{11} = \frac{e^{-2\xi}}{d\xi^2}\,
  \frac{\Sigma(\Sigma+2r)\,\partial_\xi A - 2A\!\left(r\,\partial_\xi \Sigma + \Sigma[\partial_\xi \Sigma + d\xi\, e^{\xi}]\right)}
  {\Sigma^2(\Sigma+2r)^2}
  - \frac{2 d\xi\, e^{-2\xi}}{d\xi^2}\,\frac{A}{\Sigma(\Sigma+2r)}, &                                                                                                                               &     \\
                                                                    & \partial_\eta h^{11} = \frac{e^{-2\xi}}{d\xi^2}\,
  \frac{\Sigma(\Sigma+2r)\,\partial_\eta A - 2A\,\partial_\eta \Sigma\,(r+\Sigma)}
  {\Sigma^2(\Sigma+2r)^2},                                          &                                                                                                                                     \\[2pt]
                                                                    & \partial_\xi h^{22} = -\frac{\partial_\xi \Sigma}{\Sigma^2}\,\frac{1}{d\eta^2},                                               &   &
  \partial_\eta h^{22} = -\frac{\partial_\eta \Sigma}{\Sigma^2}\,\frac{1}{d\eta^2},                                                                                                                       \\
                                                                    & \partial_\xi h^{33} = -\frac{\partial_\xi \Sigma}{\Sigma^2 \sin^2\theta},                                                     &   &
  \partial_\eta h^{33} = -\frac{\partial_\eta \Sigma + 2\frac{\cos\theta}{\sin\theta}\,\Sigma\, d\eta\,\tfrac{d\theta}{d\eta}}{\Sigma^2 \sin^2\theta},                                                    \\
                                                                    & \partial_\xi h^{13} = -\,\frac{a}{\Sigma^2}\,\partial_\xi \Sigma \,\frac{e^{-\xi}}{d\xi}
  - d\xi\,\frac{e^{-\xi}}{d\xi}\,\frac{a}{\Sigma},                  &                                                                                                                               &
  \partial_\eta h^{13} = -\,\frac{a}{\Sigma^2}\,\partial_\eta \Sigma \,\frac{e^{-\xi}}{d\xi},                                                                                                             \\[2pt]
                                                                    & \partial_\xi \Sigma = 2r\, d\xi\, e^{\xi},                                                                                    &   &
  \partial_\eta \Sigma = -2a^2 \sin\theta \cos\theta \; d\eta\,\tfrac{d\theta}{d\eta},                                                                                                                    \\
                                                                    & \partial_\xi \Delta = 2(r-1)\, d\xi\, e^{\xi},                                                                                &   &
  \partial_\eta \Delta = 0,                                                                                                                                                                               \\
                                                                    & \partial_\xi A = 4r\, d\xi\, e^{\xi}(r^2+a^2) - a^2 \sin^2\theta\, \partial_\xi \Delta,                                       &   &
  \partial_\eta A = -2a^2 \sin\theta \cos\theta\,\Delta \, d\eta\,\tfrac{d\theta}{d\eta},                                                                                                                 \\
                                                                    & \partial_\xi \alpha = -\frac{(d\xi\, e^{\xi})\,\Sigma - r\,\partial_\xi \Sigma}{\Sigma^2}\,\alpha^3,                          &   &
  \partial_\eta \alpha = \frac{r\,\partial_\eta \Sigma}{\Sigma^2}\,\alpha^3,                                                                                                                              \\
                                                                    & \partial_\xi \beta^1 = e^{-\xi} d\xi^{-1}\,\frac{2\bigl[(d\xi\, e^{\xi})\Sigma - r\,\partial_\xi \Sigma\bigr]}{(\Sigma+2r)^2}
  - d\xi\, e^{-\xi} d\xi^{-1}\,\frac{z}{1+z},                       &                                                                                                                               &
  \partial_\eta \beta^1 = -\,e^{-\xi} d\xi^{-1}\,\frac{2 r\,\partial_\eta \Sigma}{\Sigma^2 (1+z)^2}.
\end{align*}

\subsection{Transformation between BL and KS}
For comparison with other works and to connect with the physically more intuitive BL frame, it is useful to write down the explicit transformation between BL and KS coordinates. If $g_{\mu\nu}$ denotes the metric in KS coordinates and $g_{\tilde{\mu}\tilde{\nu}}$ the metric in BL coordinates, the Jacobian matrices relating the two are
\begin{equation}
  J_{\ \tilde{\nu}}^{\mu}=
  \begin{bmatrix}
    1     & 2r/\Delta & \cdot & \cdot \\
    \cdot & 1         & \cdot & \cdot \\
    \cdot & \cdot     & 1     & \cdot \\
    \cdot & a/\Delta  & \cdot & 1
  \end{bmatrix},
  \qquad
  J_{\ \nu}^{\tilde{\mu}}=
  \begin{bmatrix}
    1     & -2r/\Delta & \cdot & \cdot \\
    \cdot & 1          & \cdot & \cdot \\
    \cdot & \cdot      & 1     & \cdot \\
    \cdot & -a/\Delta  & \cdot & 1
  \end{bmatrix}.
\end{equation}
The transformations of contravariant and covariant components follow directly:
\[
  x^{\tilde{\mu}}=J^{\tilde{\mu}}_{\ \nu} x^\nu, \qquad
  x^{\mu}=J^{\mu}_{\ \tilde{\nu}} x^{\tilde{\nu}}, \qquad
  x_{\tilde{\mu}}=J_{\tilde{\mu}}^{\ \nu} x_{\nu}, \qquad
  x_{\mu}=J_{\mu}^{\ \tilde{\nu}} x_{\tilde{\nu}}.
\]

\bibliography{references, numeric}{}

@ARTICLE{Yee_1966,
       author = {{Yee}, Kane},
        title = "{Numerical solution of initial boundary value problems involving maxwell's equations in isotropic media}",
      journal = {IEEE Transactions on Antennas and Propagation},
         year = 1966,
        month = may,
       volume = {14},
       number = {3},
        pages = {302-307},
          doi = {10.1109/TAP.1966.1138693},
       adsurl = {https://ui.adsabs.harvard.edu/abs/1966ITAP...14..302Y}
}

@ARTICLE{Zou2025PhRvD.111h3032Z,
       author = {{Zou}, Minghao and {Hakobyan}, Hayk and {Mbarek}, Rostom and {Ripperda}, Bart and {Bacchini}, Fabio and {Sironi}, Lorenzo},
        title = "{New particle pusher with hadronic interactions for modeling multimessenger emission from compact objects}",
      journal = {\prd},
         year = 2025,
        month = apr,
       volume = {111},
       number = {8},
          eid = {083032},
        pages = {083032},
          doi = {10.1103/PhysRevD.111.083032},
archivePrefix = {arXiv},
       eprint = {2410.22781},
 primaryClass = {astro-ph.HE},
       adsurl = {https://ui.adsabs.harvard.edu/abs/2025PhRvD.111h3032Z}
}

@article{Komissarov_2004,
    author = {Komissarov, S. S.},
    title = {Electrodynamics of black hole magnetospheres},
    journal = {Monthly Notices of the Royal Astronomical Society},
    volume = {350},
    number = {2},
    pages = {427-448},
    year = {2004},
    month = {05},
    issn = {0035-8711},
    doi = {10.1111/j.1365-2966.2004.07598.x},
    url = {https://doi.org/10.1111/j.1365-2966.2004.07598.x},
    eprint = {https://academic.oup.com/mnras/article-pdf/350/2/427/3884469/350-2-427.pdf},
}

@ARTICLE{ThorneMacDonald1982,
       author = {{Thorne}, K.~S. and {MacDonald}, D.},
        title = "{Electrodynamics in Curved Spacetime - 3+1 Formulation}",
      journal = {\mnras},
         year = 1982,
        month = jan,
       volume = {198},
        pages = {339},
          doi = {10.1093/mnras/198.2.339},
       adsurl = {https://ui.adsabs.harvard.edu/abs/1982MNRAS.198..339T}
}

@ARTICLE{DodinFisch2010PhPl,
       author = {{Dodin}, I.~Y. and {Fisch}, N.~J.},
        title = "{Vlasov equation and collisionless hydrodynamics adapted to curved spacetime}",
      journal = {Physics of Plasmas},
         year = 2010,
        month = nov,
       volume = {17},
       number = {11},
          eid = {112118},
        pages = {112118},
          doi = {10.1063/1.3497005},
archivePrefix = {arXiv},
       eprint = {1006.3717},
 primaryClass = {physics.plasm-ph},
       adsurl = {https://ui.adsabs.harvard.edu/abs/2010PhPl...17k2118D}
}

@BOOK{Gourgoulhon2012LNP,
       author = {{Gourgoulhon}, Eric},
        title = "{3+1 Formalism in General Relativity}",
         year = 2012,
       volume = {846},
          doi = {10.1007/978-3-642-24525-1},
       adsurl = {https://ui.adsabs.harvard.edu/abs/2012LNP...846.....G}
}

@ARTICLE{ChargedPrtcl2015CQGra,
       author = {{Kolo{\v{s}}}, Martin and {Stuchl{\'\i}k}, Zden{\v{e}}k and {Tursunov}, Arman},
        title = "{Quasi-harmonic oscillatory motion of charged particles around a Schwarzschild black hole immersed in a uniform magnetic field}",
      journal = {Classical and Quantum Gravity},
         year = 2015,
        month = aug,
       volume = {32},
       number = {16},
          eid = {165009},
        pages = {165009},
          doi = {10.1088/0264-9381/32/16/165009},
archivePrefix = {arXiv},
       eprint = {1506.06799},
 primaryClass = {gr-qc},
       adsurl = {https://ui.adsabs.harvard.edu/abs/2015CQGra..32p5009K}
}

@ARTICLE{Levin_LeafOrbits2008PhRvD,
       author = {{Levin}, Janna and {Perez-Giz}, Gabe},
        title = "{A periodic table for black hole orbits}",
      journal = {\prd},
         year = 2008,
        month = may,
       volume = {77},
       number = {10},
          eid = {103005},
        pages = {103005},
          doi = {10.1103/PhysRevD.77.103005},
archivePrefix = {arXiv},
       eprint = {0802.0459},
 primaryClass = {gr-qc},
       adsurl = {https://ui.adsabs.harvard.edu/abs/2008PhRvD..77j3005L}
}

@ARTICLE{Bacchini2019ApJS,
       author = {{Bacchini}, F. and {Ripperda}, B. and {Porth}, O. and {Sironi}, L.},
        title = "{Generalized, Energy-conserving Numerical Simulations of Particles in General Relativity. II. Test Particles in Electromagnetic Fields and GRMHD}",
      journal = {\apjs},
         year = 2019,
        month = feb,
       volume = {240},
       number = {2},
          eid = {40},
        pages = {40},
          doi = {10.3847/1538-4365/aafcb3},
archivePrefix = {arXiv},
       eprint = {1810.00842},
 primaryClass = {astro-ph.HE},
       adsurl = {https://ui.adsabs.harvard.edu/abs/2019ApJS..240...40B}
}

@ARTICLE{Parfrey2019PhRvL,
       author = {{Parfrey}, Kyle and {Philippov}, Alexander and {Cerutti}, Beno{\^\i}t},
        title = "{First-Principles Plasma Simulations of Black-Hole Jet Launching}",
      journal = {\prl},
         year = 2019,
        month = jan,
       volume = {122},
       number = {3},
          eid = {035101},
        pages = {035101},
          doi = {10.1103/PhysRevLett.122.035101},
archivePrefix = {arXiv},
       eprint = {1810.03613},
 primaryClass = {astro-ph.HE},
       adsurl = {https://ui.adsabs.harvard.edu/abs/2019PhRvL.122c5101P}
}

@ARTICLE{Wald1974PhRvD..10.1680W,
       author = {{Wald}, Robert M.},
        title = "{Black hole in a uniform magnetic field}",
      journal = {\prd},
         year = 1974,
        month = sep,
       volume = {10},
       number = {6},
        pages = {1680-1685},
          doi = {10.1103/PhysRevD.10.1680},
       adsurl = {https://ui.adsabs.harvard.edu/abs/1974PhRvD..10.1680W}
}

@ARTICLE{Stuchlik2016EPJC...76...32S,
       author = {{Stuchl{\'\i}k}, Zden{\v{e}}k and {Kolo{\v{s}}}, Martin},
        title = "{Acceleration of the charged particles due to chaotic scattering in the combined black hole gravitational field and asymptotically uniform magnetic field}",
      journal = {European Physical Journal C},
         year = 2016,
        month = jan,
       volume = {76},
          eid = {32},
        pages = {32},
          doi = {10.1140/epjc/s10052-015-3862-2},
       adsurl = {https://ui.adsabs.harvard.edu/abs/2016EPJC...76...32S}
}

@ARTICLE{Komissarov2001MNRAS.326L..41K,
       author = {{Komissarov}, S.~S.},
        title = "{Direct numerical simulations of the Blandford-Znajek effect}",
      journal = {\mnras},
         year = 2001,
        month = sep,
       volume = {326},
       number = {3},
        pages = {L41-L44},
          doi = {10.1046/j.1365-8711.2001.04863.x},
       adsurl = {https://ui.adsabs.harvard.edu/abs/2001MNRAS.326L..41K}
}

@ARTICLE{KomissarovMonopole2004MNRAS.350.1431K,
       author = {{Komissarov}, S.~S.},
        title = "{General relativistic magnetohydrodynamic simulations of monopole magnetospheres of black holes}",
      journal = {\mnras},
         year = 2004,
        month = jun,
       volume = {350},
       number = {4},
        pages = {1431-1436},
          doi = {10.1111/j.1365-2966.2004.07738.x},
archivePrefix = {arXiv},
       eprint = {astro-ph/0402430},
 primaryClass = {astro-ph},
       adsurl = {https://ui.adsabs.harvard.edu/abs/2004MNRAS.350.1431K}
}

@ARTICLE{BlandfordZnajek1977MNRAS.179..433B,
       author = {{Blandford}, R.~D. and {Znajek}, R.~L.},
        title = "{Electromagnetic extraction of energy from Kerr black holes.}",
      journal = {\mnras},
         year = 1977,
        month = may,
       volume = {179},
        pages = {433-456},
          doi = {10.1093/mnras/179.3.433},
       adsurl = {https://ui.adsabs.harvard.edu/abs/1977MNRAS.179..433B}
}

@ARTICLE{APERTURE,
       author = {{Chen}, Alexander Y. and {Luepker}, Martin and {Yuan}, Yajie},
        title = "{Introducing APERTURE: A GPU-based General Relativistic Particle-in-Cell Simulation Framework}",
      journal = {arXiv e-prints},
         year = 2025,
        month = mar,
          eid = {arXiv:2503.04558},
        pages = {arXiv:2503.04558},
          doi = {10.48550/arXiv.2503.04558},
archivePrefix = {arXiv},
       eprint = {2503.04558},
 primaryClass = {astro-ph.HE},
       adsurl = {https://ui.adsabs.harvard.edu/abs/2025arXiv250304558C}
}

@ARTICLE{EHTcode,
       author = {{Porth}, Oliver and {Chatterjee}, Koushik and {Narayan}, Ramesh and {Gammie}, Charles F. and {Mizuno}, Yosuke and {Anninos}, Peter and {Baker}, John G. and {Bugli}, Matteo and {Chan}, Chi-kwan and {Davelaar}, Jordy and {Del Zanna}, Luca and {Etienne}, Zachariah B. and {Fragile}, P. Chris and {Kelly}, Bernard J. and {Liska}, Matthew and {Markoff}, Sera and {McKinney}, Jonathan C. and {Mishra}, Bhupendra and {Noble}, Scott C. and {Olivares}, H{\'e}ctor and {Prather}, Ben and {Rezzolla}, Luciano and {Ryan}, Benjamin R. and {Stone}, James M. and {Tomei}, Niccol{\`o} and {White}, Christopher J. and {Younsi}, Ziri and {Akiyama}, Kazunori and {Alberdi}, Antxon and {Alef}, Walter and {Asada}, Keiichi and {Azulay}, Rebecca and {Baczko}, Anne-Kathrin and {Ball}, David and {Balokovi{\'c}}, Mislav and {Barrett}, John and {Bintley}, Dan and {Blackburn}, Lindy and {Boland}, Wilfred and {Bouman}, Katherine L. and {Bower}, Geoffrey C. and {Bremer}, Michael and {Brinkerink}, Christiaan D. and {Brissenden}, Roger and {Britzen}, Silke and {Broderick}, Avery E. and {Broguiere}, Dominique and {Bronzwaer}, Thomas and {Byun}, Do-Young and {Carlstrom}, John E. and {Chael}, Andrew and {Chatterjee}, Shami and {Chen}, Ming-Tang and {Chen}, Yongjun and {Cho}, Ilje and {Christian}, Pierre and {Conway}, John E. and {Cordes}, James M. and {Geoffrey} and {Crew}, B. and {Cui}, Yuzhu and {De Laurentis}, Mariafelicia and {Deane}, Roger and {Dempsey}, Jessica and {Desvignes}, Gregory and {Doeleman}, Sheperd S. and {Eatough}, Ralph P. and {Falcke}, Heino and {Fish}, Vincent L. and {Fomalont}, Ed and {Fraga-Encinas}, Raquel and {Freeman}, Bill and {Friberg}, Per and {Fromm}, Christian M. and {G{\'o}mez}, Jos{\'e} L. and {Galison}, Peter and {Garc{\'\i}a}, Roberto and {Gentaz}, Olivier and {Georgiev}, Boris and {Goddi}, Ciriaco and {Gold}, Roman and {Gu}, Minfeng and {Gurwell}, Mark and {Hada}, Kazuhiro and {Hecht}, Michael H. and {Hesper}, Ronald and {Ho}, Luis C. and {Ho}, Paul and {Honma}, Mareki and {Huang}, Chih-Wei L. and {Huang}, Lei and {Hughes}, David H. and {Ikeda}, Shiro and {Inoue}, Makoto and {Issaoun}, Sara and {James}, David J. and {Jannuzi}, Buell T. and {Janssen}, Michael and {Jeter}, Britton and {Jiang}, Wu and {Johnson}, Michael D. and {Jorstad}, Svetlana and {Jung}, Taehyun and {Karami}, Mansour and {Karuppusamy}, Ramesh and {Kawashima}, Tomohisa and {Keating}, Garrett K. and {Kettenis}, Mark and {Kim}, Jae-Young and {Kim}, Junhan and {Kim}, Jongsoo and {Kino}, Motoki and {Koay}, Jun Yi and {Patrick} and {Koch}, M. and {Koyama}, Shoko and {Kramer}, Michael and {Kramer}, Carsten and {Krichbaum}, Thomas P. and {Kuo}, Cheng-Yu and {Lauer}, Tod R. and {Lee}, Sang-Sung and {Li}, Yan-Rong and {Li}, Zhiyuan and {Lindqvist}, Michael and {Liu}, Kuo and {Liuzzo}, Elisabetta and {Lo}, Wen-Ping and {Lobanov}, Andrei P. and {Loinard}, Laurent and {Lonsdale}, Colin and {Lu}, Ru-Sen and {MacDonald}, Nicholas R. and {Mao}, Jirong and {Marrone}, Daniel P. and {Marscher}, Alan P. and {Mart{\'\i}-Vidal}, Iv{\'a}n and {Matsushita}, Satoki and {Matthews}, Lynn D. and {Medeiros}, Lia and {Menten}, Karl M. and {Mizuno}, Izumi and {Moran}, James M. and {Moriyama}, Kotaro and {Moscibrodzka}, Monika and {M{\"u}ller}, Cornelia and {Nagai}, Hiroshi and {Nagar}, Neil M. and {Nakamura}, Masanori and {Narayanan}, Gopal and {Natarajan}, Iniyan and {Neri}, Roberto and {Ni}, Chunchong and {Noutsos}, Aristeidis and {Okino}, Hiroki and {Oyama}, Tomoaki and {{\"O}zel}, Feryal and {Palumbo}, Daniel C.~M. and {Patel}, Nimesh and {Pen}, Ue-Li and {Pesce}, Dominic W. and {Pi{\'e}tu}, Vincent and {Plambeck}, Richard and {PopStefanija}, Aleksandar and {Preciado-L{\'o}pez}, Jorge A. and {Psaltis}, Dimitrios and {Pu}, Hung-Yi and {Ramakrishnan}, Venkatessh and {Rao}, Ramprasad and {Rawlings}, Mark G. and {Raymond}, Alexander W. and {Ripperda}, Bart and {Roelofs}, Freek and {Rogers}, Alan and {Ros}, Eduardo and {Rose}, Mel and {Roshanineshat}, Arash and {Rottmann}, Helge and {Roy}, Alan L. and {Ruszczyk}, Chet and {Rygl}, Kazi L.~J. and {S{\'a}nchez}, Salvador and {S{\'a}nchez-Arguelles}, David and {Sasada}, Mahito and {Savolainen}, Tuomas and {Schloerb}, F. Peter and {Schuster}, Karl-Friedrich and {Shao}, Lijing and {Shen}, Zhiqiang and {Small}, Des and {Sohn}, Bong Won and {SooHoo}, Jason and {Tazaki}, Fumie and {Tiede}, Paul and {Tilanus}, Remo P.~J. and {Titus}, Michael and {Toma}, Kenji and {Torne}, Pablo and {Trent}, Tyler and {Trippe}, Sascha and {Tsuda}, Shuichiro and {van Bemmel}, Ilse and {van Langevelde}, Huib Jan and {van Rossum}, Daniel R. and {Wagner}, Jan and {Wardle}, John and {Weintroub}, Jonathan and {Wex}, Norbert and {Wharton}, Robert and {Wielgus}, Maciek and {Wong}, George N. and {Wu}, Qingwen and {Young}, Ken and {Young}, Andr{\'e} and {Yuan}, Feng and {Yuan}, Ye-Fei and {Zensus}, J. Anton and {Zhao}, Guangyao and {Zhao}, Shan-Shan and {Zhu}, Ziyan and {Event Horizon Telescope Collaboration}},
        title = "{The Event Horizon General Relativistic Magnetohydrodynamic Code Comparison Project}",
      journal = {\apjs},
         year = 2019,
        month = aug,
       volume = {243},
       number = {2},
          eid = {26},
        pages = {26},
          doi = {10.3847/1538-4365/ab29fd},
archivePrefix = {arXiv},
       eprint = {1904.04923},
 primaryClass = {astro-ph.HE},
       adsurl = {https://ui.adsabs.harvard.edu/abs/2019ApJS..243...26P}
}

@ARTICLE{YuanNarayan2014ARA&A..52..529Y,
       author = {{Yuan}, Feng and {Narayan}, Ramesh},
        title = "{Hot Accretion Flows Around Black Holes}",
      journal = {\araa},
         year = 2014,
        month = aug,
       volume = {52},
        pages = {529-588},
          doi = {10.1146/annurev-astro-082812-141003},
archivePrefix = {arXiv},
       eprint = {1401.0586},
 primaryClass = {astro-ph.HE},
       adsurl = {https://ui.adsabs.harvard.edu/abs/2014ARA&A..52..529Y}
}

@ARTICLE{FPIC2025ApJ...992L...8M,
       author = {{Meringolo}, Claudio and {Camilloni}, Filippo and {Rezzolla}, Luciano},
        title = "{Electromagnetic Energy Extraction from Kerr Black Holes: Ab Initio Calculations}",
      journal = {\apjl},
         year = 2025,
        month = oct,
       volume = {992},
       number = {1},
          eid = {L8},
        pages = {L8},
          doi = {10.3847/2041-8213/ae06a6},
archivePrefix = {arXiv},
       eprint = {2507.08942},
 primaryClass = {gr-qc},
       adsurl = {https://ui.adsabs.harvard.edu/abs/2025ApJ...992L...8M}
}

@BOOK{Weinberg_1972,
       author = {{Weinberg}, Steven},
        title = "{Gravitation and Cosmology: Principles and Applications of the General Theory of Relativity}",
         year = 1972,
       adsurl = {https://ui.adsabs.harvard.edu/abs/1972gcpa.book.....W}
}

@ARTICLE{Galeev1979ApJ...229..318G,
       author = {{Galeev}, A.~A. and {Rosner}, R. and {Vaiana}, G.~S.},
        title = "{Structured coronae of accretion disks.}",
      journal = {\apj},
         year = 1979,
        month = apr,
       volume = {229},
        pages = {318-326},
          doi = {10.1086/156957},
       adsurl = {https://ui.adsabs.harvard.edu/abs/1979ApJ...229..318G}
}

@ARTICLE{Beloborodov2017ApJ...850..141B,
       author = {{Beloborodov}, Andrei M.},
        title = "{Radiative Magnetic Reconnection Near Accreting Black Holes}",
      journal = {\apj},
         year = 2017,
        month = dec,
       volume = {850},
       number = {2},
          eid = {141},
        pages = {141},
          doi = {10.3847/1538-4357/aa8f4f},
archivePrefix = {arXiv},
       eprint = {1701.02847},
 primaryClass = {astro-ph.HE},
       adsurl = {https://ui.adsabs.harvard.edu/abs/2017ApJ...850..141B}
}

@ARTICLE{Bambic2024MNRAS.527.2895B,
       author = {{Bambic}, Christopher J. and {Quataert}, Eliot and {Kunz}, Matthew W.},
        title = "{Local models of two-temperature accretion disc coronae - I. Structure, outflows, and energetics}",
      journal = {\mnras},
         year = 2024,
        month = jan,
       volume = {527},
       number = {2},
        pages = {2895-2918},
          doi = {10.1093/mnras/stad3261},
archivePrefix = {arXiv},
       eprint = {2304.06067},
 primaryClass = {astro-ph.HE},
       adsurl = {https://ui.adsabs.harvard.edu/abs/2024MNRAS.527.2895B}
}

@ARTICLE{GRAVITY2018,
       author = {{GRAVITY Collaboration} and {Abuter}, R. and {Amorim}, A. and {Baub{\"o}ck}, M. and {Berger}, J.~P. and {Bonnet}, H. and {Brandner}, W. and {Cl{\'e}net}, Y. and {Coud{\'e} Du Foresto}, V. and {de Zeeuw}, P.~T. and {Deen}, C. and {Dexter}, J. and {Duvert}, G. and {Eckart}, A. and {Eisenhauer}, F. and {F{\"o}rster Schreiber}, N.~M. and {Garcia}, P. and {Gao}, F. and {Gendron}, E. and {Genzel}, R. and {Gillessen}, S. and {Guajardo}, P. and {Habibi}, M. and {Haubois}, X. and {Henning}, Th. and {Hippler}, S. and {Horrobin}, M. and {Huber}, A. and {Jim{\'e}nez-Rosales}, A. and {Jocou}, L. and {Kervella}, P. and {Lacour}, S. and {Lapeyr{\`e}re}, V. and {Lazareff}, B. and {Le Bouquin}, J. -B. and {L{\'e}na}, P. and {Lippa}, M. and {Ott}, T. and {Panduro}, J. and {Paumard}, T. and {Perraut}, K. and {Perrin}, G. and {Pfuhl}, O. and {Plewa}, P.~M. and {Rabien}, S. and {Rodr{\'\i}guez-Coira}, G. and {Rousset}, G. and {Sternberg}, A. and {Straub}, O. and {Straubmeier}, C. and {Sturm}, E. and {Tacconi}, L.~J. and {Vincent}, F. and {von Fellenberg}, S. and {Waisberg}, I. and {Widmann}, F. and {Wieprecht}, E. and {Wiezorrek}, E. and {Woillez}, J. and {Yazici}, S.},
        title = "{Detection of orbital motions near the last stable circular orbit of the massive black hole SgrA*}",
      journal = {\aap},
         year = 2018,
        month = oct,
       volume = {618},
          eid = {L10},
        pages = {L10},
          doi = {10.1051/0004-6361/201834294},
archivePrefix = {arXiv},
       eprint = {1810.12641},
 primaryClass = {astro-ph.GA},
       adsurl = {https://ui.adsabs.harvard.edu/abs/2018A&A...618L..10G}
}

@ARTICLE{Dodds-Eden2009ApJ...698..676D,
       author = {{Dodds-Eden}, K. and {Porquet}, D. and {Trap}, G. and {Quataert}, E. and {Haubois}, X. and {Gillessen}, S. and {Grosso}, N. and {Pantin}, E. and {Falcke}, H. and {Rouan}, D. and {Genzel}, R. and {Hasinger}, G. and {Goldwurm}, A. and {Yusef-Zadeh}, F. and {Clenet}, Y. and {Trippe}, S. and {Lagage}, P. -O. and {Bartko}, H. and {Eisenhauer}, F. and {Ott}, T. and {Paumard}, T. and {Perrin}, G. and {Yuan}, F. and {Fritz}, T.~K. and {Mascetti}, L.},
        title = "{Evidence for X-Ray Synchrotron Emission from Simultaneous Mid-Infrared to X-Ray Observations of a Strong Sgr A* Flare}",
      journal = {\apj},
         year = 2009,
        month = jun,
       volume = {698},
       number = {1},
        pages = {676-692},
          doi = {10.1088/0004-637X/698/1/676},
archivePrefix = {arXiv},
       eprint = {0903.3416},
 primaryClass = {astro-ph.GA},
       adsurl = {https://ui.adsabs.harvard.edu/abs/2009ApJ...698..676D}
}

@ARTICLE{Yusef-Zadeh2009ApJ,
       author = {{Yusef-Zadeh}, F. and {Bushouse}, H. and {Wardle}, M. and {Heinke}, C. and {Roberts}, D.~A. and {Dowell}, C.~D. and {Brunthaler}, A. and {Reid}, M.~J. and {Martin}, C.~L. and {Marrone}, D.~P. and {Porquet}, D. and {Grosso}, N. and {Dodds-Eden}, K. and {Bower}, G.~C. and {Wiesemeyer}, H. and {Miyazaki}, A. and {Pal}, S. and {Gillessen}, S. and {Goldwurm}, A. and {Trap}, G. and {Maness}, H.},
        title = "{Simultaneous Multi-Wavelength Observations of Sgr A* During 2007 April 1-11}",
      journal = {\apj},
         year = 2009,
        month = nov,
       volume = {706},
       number = {1},
        pages = {348-375},
          doi = {10.1088/0004-637X/706/1/348},
archivePrefix = {arXiv},
       eprint = {0907.3786},
 primaryClass = {astro-ph.GA},
       adsurl = {https://ui.adsabs.harvard.edu/abs/2009ApJ...706..348Y}
}

@ARTICLE{Fellenberg2025ApJ...979L..20V,
       author = {{von Fellenberg}, Sebastiano D. and {Roychowdhury}, Tamojeet and {Michail}, Joseph M. and {Sumners}, Zach and {Sanger-Johnson}, Grace and {Fazio}, Giovanni G. and {Haggard}, Daryl and {Hora}, Joseph L. and {Philippov}, Alexander and {Ripperda}, Bart and {Smith}, Howard A. and {Willner}, S.~P. and {Witzel}, Gunther and {Zhang}, Shuo and {Becklin}, Eric E. and {Bower}, Geoffrey C. and {Chandra}, Sunil and {Do}, Tuan and {Garcia Marin}, Macarena and {Gurwell}, Mark A. and {Ford}, Nicole M. and {Hada}, Kazuhiro and {Markoff}, Sera and {Morris}, Mark R. and {Neilsen}, Joey and {Sabha}, Nadeen B. and {Seefeldt-Gail}, Braden},
        title = "{First Mid-infrared Detection and Modeling of a Flare from Sgr A*}",
      journal = {\apjl},
         year = 2025,
        month = jan,
       volume = {979},
       number = {1},
          eid = {L20},
        pages = {L20},
          doi = {10.3847/2041-8213/ada3d2},
archivePrefix = {arXiv},
       eprint = {2501.07415},
 primaryClass = {astro-ph.HE},
       adsurl = {https://ui.adsabs.harvard.edu/abs/2025ApJ...979L..20V}
}
\bibliographystyle{aasjournalv7}

\end{document}